\documentclass[lettersize,journal,final,twoside,twocolumn]{IEEEtran}

\usepackage[utf8]{inputenc}
\usepackage{amsmath, bbm}
\usepackage{amssymb}
\usepackage{amsfonts}
\usepackage{amsthm}
\usepackage{algorithm}
\usepackage{algorithmic,setspace}
\usepackage{nicefrac}
\usepackage{bm}
\usepackage{hyperref}
\usepackage{cite}
\usepackage[T1]{fontenc}
\usepackage{graphicx}
\usepackage[font=small]{caption}
\usepackage{tikz}
\usetikzlibrary{positioning}
\usetikzlibrary{arrows}
\usetikzlibrary{matrix}
\usetikzlibrary{backgrounds}
\usepackage{pgfplots}
\usepackage{epstopdf}
\usepackage[caption=false,font=footnotesize,labelformat=simple]{subfig}
\usepackage[export]{adjustbox}
\usepackage{multirow}
\usepackage{lipsum}
\usepackage{mathtools}

\IEEEoverridecommandlockouts
\usepackage{etoolbox}
\makeatletter
\patchcmd{\@makecaption}
{\scshape}
{}
{}
{}
\makeatletter
\patchcmd{\@makecaption}
{\\}
{.\ }
{}
{}
\newtheorem{Problem}{Problem}

\newlength\imageheight
\input{mysymbol.sty}

\newcommand \COMMENTS[1] {{\color{gray}\textit{/* #1 */}}} 

\title{Link Scheduling using Graph Neural Networks}

\author{\IEEEauthorblockN{Zhongyuan Zhao,~\IEEEmembership{Member,~IEEE,} 
        Gunjan Verma,
        Chirag Rao,
        Ananthram Swami,~\IEEEmembership{Life Fellow,~IEEE,}\\ 
        and Santiago Segarra,~\IEEEmembership{Senior Member,~IEEE}}
	\thanks{Z. Zhao and S. Segarra are with the Department of Electrical and Computer Engineering, Rice University, USA. e-mails: \{zhongyuan.zhao, segarra\}@rice.edu}
	\thanks{G. Verma, C. Rao, and A. Swami are with the US Army’s DEVCOM Army Research Laboratory, USA. e-mails: \{gunjan.verma.civ, chirag.r.rao.civ, ananthram.swami.civ\}@army.mil.}
	\thanks{Research was sponsored by the Army Research Office and was accomplished under Cooperative Agreement Number W911NF-19-2-0269. 
		The views and conclusions contained in this document are those of the authors and should not be interpreted as representing the official policies, either expressed or implied, of the Army Research Office or the U.S. Government. 
		The U.S. Government is authorized to reproduce and distribute reprints for Government purposes notwithstanding any copyright notation herein.}
	\thanks{Preliminary results were presented in~\cite{zhao2021icassp}.}
}

\pgfplotsset{compat=1.16}
\begin{document}
	
\markboth{IEEE Transactions on Wireless Communications, ,~Vol.~xx, No.~x, XXX~2023}{Zhao \MakeLowercase{\textit{et al.}}: Link Scheduling using Graph Neural Networks}

\IEEEpubid{0000--0000/00\$00.00~\copyright~2023 IEEE}

\maketitle


\begin{abstract}%
Efficient scheduling of transmissions is a key problem in wireless networks. 
The main challenge stems from the fact that optimal link scheduling involves solving a maximum weighted independent set (MWIS) problem, which is known to be NP-hard. 
In practical schedulers, centralized and distributed greedy heuristics are commonly used to approximately solve the MWIS problem.
However, most of these greedy heuristics ignore important topological information of the wireless network.
To overcome this limitation, we propose fast heuristics based on graph convolutional networks (GCNs) that can be implemented in centralized and distributed manners.
Our centralized heuristic is based on tree search guided by a GCN and 1-step rollout.
In our distributed MWIS solver, a GCN generates topology-aware node embeddings that are combined with per-link utilities before invoking a distributed greedy solver. 
Moreover, a novel reinforcement learning scheme is developed to train the GCN in a non-differentiable pipeline.
Test results on medium-sized wireless networks show that our centralized heuristic can reach a near-optimal solution quickly, and our distributed heuristic based on a shallow GCN can reduce by nearly half the suboptimality gap of the distributed greedy solver with minimal increase in complexity. 
The proposed schedulers also exhibit good generalizability across graph and weight distributions.
\end{abstract}


\section{Introduction}\label{sec:intro}

A fundamental problem in managing wireless networks is to efficiently schedule transmissions. 
In general, the scheduling problem involves determining which links should transmit and when they should transmit, along with other relevant parameters such as transmit power, modulation, and coding schemes~\cite{tassiulas1992,Joo09,marques2011optimal}. 
In this paper, we focus on link scheduling in wireless networks with time-slotted orthogonal multiple access, in which a time slot comprises a scheduling phase followed by a transmission phase \cite{Kabbani07,paschalidis2015message}. 
This problem is associated with many real-world applications of both ad-hoc and infrastructure-based wireless networks, e.g., battlefield communications, vehicular/flying ad-hoc networks, wireless sensor networks~\cite{Lin06,sarkar2013ad}, device-to-device (D2D) communications~\cite{Kim2016d2d}, cloud-radio access networks (CRAN) \cite{Douik2018}, wireless backhaul networks~\cite{Kabbani07,gupta2020learning}, multihop relay networks for mmWave and THz bands communications~\cite{qiao2011enabling,xia2017cross}, and internet of things~\cite{kott2016internet,akyildiz20206g}.
Compared to contention-based multiple access, e.g. CSMA-CA, scheduled multiple access is often preferred due to its increased spectrum utilization efficiency~\cite{Jindal13}.
This benefit, however, comes with the associated challenge that optimal scheduling involves solving a maximum weighted independent set (MWIS) problem~\cite{zhao2021icassp,tassiulas1992,Kabbani07,Joo09,joo2012local,joo2015distributed,marques2011optimal,sanghavi2009message,du2016new,Li18Ising,Douik2018,paschalidis2015message,joo2010complexity}. 
An MWIS is the independent set (IS) that achieves maximal total weight on a node-weighted graph, where an IS is a set of vertices not connected by any edges.
The MWIS problem is NP-hard -- the running time of exactly solving it grows as an exponential function of the size of the graph~\cite{cheng2009complexity,joo2010complexity}.
The idea of scheduling links by selecting the MWIS of the conflict graph of a network was first established in \cite{tassiulas1992}, and has been developed since then~\cite{Kabbani07,Joo09,joo2012local,joo2015distributed,marques2011optimal,sanghavi2009message,du2016new,Li18Ising,Douik2018,paschalidis2015message,joo2010complexity}. 
Such link scheduling schemes involve two major tasks: 
1) A per-link utility function to compute the weights associated with activating each link, and
2) An efficient (possibly distributed) \textit{MaxWeight} {\cite{tassiulas1992,joo2015distributed}} scheduler as an approximate solver for the associated MWIS problem. 
\IEEEpubidadjcol

In terms of the per-link utility design, queue length~\cite{Joo09,joo2012local,tassiulas1992}, link rate~\cite{Douik2018}, their product~\cite{joo2015distributed} and ratio~\cite{paschalidis2015message}, and the age of information~\cite{kadota2018scheduling} have been used in the past, along with some more theoretically-grounded variations~\cite{marques2011optimal}. 
Moreover, routing decisions can be involved in the per-link utility. 
In backpressure routing \cite{tassiulas1992,xue2013delay}, the per-link utility for scheduling is the queue length of a flow decided by a routing algorithm for throughput-optimal routing \cite{tassiulas1992} or a virtual queue length for delay-optimal routing \cite{xue2013delay}.
Utility functions can promote not only traffic metrics, but also fairness across the network \cite{ogryczak2014fair}.
Besides analytical functions, the per-link utilities can also be generated by a reinforcement learning (RL) agent with a continuous action space \cite{gupta2020learning}, where the transition of network state is modeled as a Markov decision process. 

For MaxWeight scheduling, a good MWIS solver is defined by high quality and low complexity, and should be agnostic to the choice of the per-link utility function. 
The quality of a solution is measured by its approximation ratio (AR), its ratio to the optimal solution.
Further, low communication and computational complexity can reduce scheduling overhead. 
Many heuristics have been developed for lower complexity and/or better performance \cite{dimakis2006sufficient,sanghavi2009message,paschalidis2015message,du2016new,Lucas14Ising,Li18Ising,Douik2018,Joo09,joo2015distributed,joo2012local}. 
Centralized MWIS solvers, with limited scalability but high AR provided by the global information of the network, can be used as schedulers for infrastructure-based networks \cite{dimakis2006sufficient,Kim2016d2d,Douik2018,Kabbani07,gupta2020learning}, and for theoretical exploration of network capacity \cite{marques2011optimal}.
On the other hand, distributed MWIS solvers are preferred for practical scheduling in wireless ad-hoc networks since they can reduce the computational overhead through parallelism, and eliminate the need for relaying packets over the network by limiting communications to local neighborhoods.
Moreover, distributed scheduling increases network robustness against a single point of failure at the fusion center.
Without global information, distributed MWIS solvers seek to achieve reasonable performance efficiently, e.g., with linear \cite{sanghavi2009message,paschalidis2015message,du2016new,Douik2018}, logarithmic \cite{Joo09,joo2015distributed,joo2012local}, or constant \cite{joo2012local,Lucas14Ising,Li18Ising} local communication complexity (referred to as \emph{local complexity}), 
which is defined as the required rounds of information exchange in the neighborhood (referred to as \emph{local exchange}).
As the common choice of practical scheduling schemes, distributed greedy heuristics~\cite{Joo09,joo2015distributed,joo2012local} seeks to mimic an iterative process of adding the link with the largest utility to the partial solution and removing other links that might interfere with it.

Motivated by the success of machine learning in other fields, data-driven learning-based solutions for resource allocation problems in wireless networks have been proposed over the last few years~\cite{lee2018deep, wang2018deep, nasir2019multi, zhang2019deep, qin2019deep,chowdhury2020unfolding, kumar2021icassp}.
A common practice in most of these approaches is to parameterize a function of interest using multi-layer perceptrons (MLP) or convolutional neural networks (CNNs), which are not well-suited for problems in wireless communications since they do not exploit the underlying topology.
This led to several approaches that tried to adjust CNNs to the wireless setting \cite{lee2018deep,xu2019energy,van2019sum,cui2019spatial}. 
Here, we adopt an alternative direction~\cite{eisen2020optimal,chowdhury2020unfolding}, in which graph neural networks (GNNs) \cite{kipf2016semi,gama2018convolutional,roddenberry_2019_hodgenet,yang_2018_enhancing} are used to incorporate the topology of the wireless network into the learning algorithm.

Our approach is also in line with the recent trend of using deep learning to find approximate solutions to combinatorial problems on graphs~\cite{khalil2017learning,li2018combinatorial}.
{More precisely, we integrate graph convolutional networks (GCNs) \cite{scarselli2008graph,battaglia2018relational,wu2020comprehensive} into fast and/or distributed algorithmic frameworks, to find solution of high quality in a timely and/or distributed manner.}
Our \emph{centralized} MWIS solver adopts 1-step lookahead rollout  \cite[Ch.~2.3]{bertsekas2021rollout} to further guide the tree search in \cite{li2018combinatorial}, to reach a good solution by traversing the search tree only once, rather than hundreds of thousands of times as in \cite{li2018combinatorial}, thus reducing the complexity by several orders of magnitude.
{Meanwhile, distributed schedulers are preferred in large networks for improved time complexity, scalability, and robustness against single-point-of-failure, since the parallelization of solution construction not only lowers the time complexity, but also eliminates the need for full knowledge of the graph.}
Our \emph{distributed} MWIS solver has a novel modular structure where a GCN-based node embedding module is followed by a distributed greedy heuristic, thus exploiting the efficiency of the latter while raising graph-awareness through the use of the GCN.
Moreover, the GCN can be trained in a computationally efficient manner -- i.e., without the need to exactly solve any MWIS problem -- and, although the training must be centralized, the execution is fully distributed. 

{\bf Contributions:} 
We develop efficient approximate MWIS solvers suitable for link scheduling in wireless networks. Specifically: 
1)~We propose the first GCN-based distributed MWIS solver for link scheduling by combining the topology-awareness of GCNs and the efficiency of distributed greedy solvers;
2)~We propose fast centralized MWIS solvers based on a GCN-guided tree search that can achieve near optimal performance on medium-sized graphs with hundreds of nodes;
3)~We develop a customized scheme of graph-based deterministic policy gradient to train the GCNs embedded in a non-differentiable downstream pipeline with the help of efficient heuristics. 
Our reinforcement learning scheme has better performance and computational efficiency than alternative supervised learning schemes,  
and complements the existing approaches of zeroth-order optimization \cite{liu2020primer} and surrogate gradient \cite{poganvcic2019differentiation} for training pipelines with blackbox/combinatorial module;
and 4) Through numerical experiments, we demonstrate the superior performance of the proposed method in single and multi-channel scheduling as well as its generalizability over different graph types and weight distributions. 

{\bf Paper outline: }The rest of this paper is organized as follows. 
Related work is reviewed in Section~\ref{sec:review}. 
The system model and the formulation of the scheduling problem are introduced in Section~\ref{sec:problem}. 
The proposed GCN-based \emph{centralized} MWIS solver is described in Section~\ref{sec:search}, followed by our GCN-based \emph{distributed} MWIS solver in Section~\ref{sec:solution}. 
Our reinforcement learning scheme is introduced in Section~\ref{sec:train}.
In Section~\ref{sec:results}, numerical experiments illustrate the performance of our proposed solutions in comparison with current state-of-the-art methods. 
Section~\ref{sec:conclusions} wraps up the paper with a short conclusion and a discussion on future directions. 

{\bf Notation: } 
The notational convention and descriptions of major notations are listed in Table~\ref{tab:symbols}.

\begin{table*}[htbp]
	\renewcommand{\arraystretch}{1.15}
	\vspace{-0.1in}
	\caption{Table of Notations 
	} 
	\vspace{-0.1in}
	\label{tab:symbols}
	\centering
	\footnotesize
	\begin{tabular}{p{1.1cm}|p{6.9cm}}
		Symbols  & Descriptions   \\ \hline
		$ (\cdot)^\top, \odot $ & $(\cdot)^\top$: transpose operator, $\odot$: element-wise product operator. \\ \hline
		$ |\cdot| $ & $|\cdot|$: cardinality of a set. \\ \hline
		$ \alpha $ & learning rate \\ \hline
		$ \gamma, \gamma(\cdot) $ & ratio of total utilities of solutions found by GCN-based solver and greedy solver, $ \gamma={u(\hat{\boldsymbol{v}}_{\mathrm{GCN}})}/{u(\hat{\boldsymbol{v}}_{\mathrm{Gr}})} $ \\ \hline
		$\bbTheta_{0}^{l},\bbTheta_{1}^{l}$ & the sets of trainable parameters of the layer $l$ of GCN in \eqref{E:gcn}  \\ \hline
		$ \lambda $ & $\lambda$: arrival rate \\ \hline
		$ \mu, \mu^{(s)} $ & $\mu=\mathbb{E}(r)/\lambda$: traffic load, \; $\mu^{(s)}$: saturation traffic load  \\ \hline
		$\sigma_{l}(\cdot)$ & activation function of layer $l$ of GCN  \\ \hline
		$\varnothing,\phi$ & $\varnothing$: an empty set (or queue),\;\;  $\phi$: an empty graph  \\ \hline
		$\bbPsi_{\ccalG}(\cdot;\bbXi)$ & parameterized function of GCN defined on graph $\ccalG$, with a set of trainable parameters $\bbXi$  \\ \hline
		$ \Omega $, \newline $ \Omega_S  $,\newline $  \Omega_{u}^{S} $ & $\Omega$: distribution of network state $(\ccalG,\bbS,\bbu)$, \newline $\Omega_S$: distribution of state $S=(\ccalG,\bbS)$, \newline  $\Omega_{u}^{S}$: conditional distribution of $\bbu$ given state $S=(\ccalG,\bbS)$ \\ \hline
		$B, b$ & $B$: branching factor of search tree, $b\in\{1,\dots,B\}$: index of a child state (branch) of current state  \\ \hline
		$\bbc$,\newline $\bbc_{v},c(v)$ &  $\bbc$: vector of control messages from all links,\newline $\bbc_{v}=c(v)$: control message from link $v$ to its neighbors \\ \hline
		$d(v), \bar{d}$ &  $d(v)$:degree of vertex $v$, $\bar{d}$:average vertex degree of graph $\ccalG$  \\ \hline
		$ \mathbb{E}(\cdot), \mathbbm{1}(\cdot) $ & $\mathbb{E}(\cdot)$: expectation,\; $\mathbbm{1}(\cdot)$: indicator function \\ \hline
		$g_{l}$,\newline $g$ & $g_{l}$: dimension of output features of layer $l$ of GCN,\newline $g$: dimension of input features of inner layers \\ \hline
		$g_{c}(\cdot)$,\newline $g_{d}(\cdot)$ &  $g_{c}(\cdot)$: function of centralized greedy heuristic,\newline $g_{d}(\cdot)$: function of distributed greedy heuristic  \\ \hline
		$\ccalG$,\newline $\ccalG(\ccalV,\ccalE)$ & Conflict graph $\ccalG$, composed of a set of vertices $\ccalV$ and a set of edges $\ccalE$. \\ \hline
		$  J(\bbXi) $ & the objective function of trainable parameters $\bbXi$ in the training formulation in Section~\ref{sec:train}  \\ \hline
		$ \nabla,\nabla\! J(\bbXi) $ & $\nabla$: gradient,\;\; $\nabla\! J(\bbXi)$ gradient of objective function $J(\bbXi)$  \\ \hline
		$k,K$,\newline $\ccalK$ & $k$: sub-channel index, $K$: number of sub-channels,\newline $\ccalK=\{1,\dots,K\}$: set of orthogonal sub-channels \\ \hline
		$L$,\newline $l$ & $L$: the total number of layers of GCN,\newline $l\in\{0,\dots,L\}$: the index of a layer  \\ \hline
		$\ccalL$ & normalized Laplacian matrix of graph $\ccalG$  \\ \hline
		$ m $ & number of edges formed by a new vertex in the preferential attachment process in Barabási–Albert model  \\ \hline
		$ N $ & maximum number of iterations in the truncated LGS-$N$ \\ \hline
	\end{tabular}
	\quad
	\begin{tabular}{p{1.5cm}|p{6.5cm}}
		Symbols  & Descriptions   \\ \hline
		$\ccalN_{\ccalG}(v),\ccalN(v)$ &  the set of immediate neighbors of vertex $v$ on graph $\ccalG$  \\ \hline
		$\mathbb{N}(a,b) $ & normal distribution with mean $a$ \& standard deviation $b$  \\ \hline
		$\ccalO(\cdot)$ & big O notation provides an upper bound on the growth rate of the function, for complexity  \\ \hline
		$ p $ & probability of edge appearance in Erdős–Rényi model  \\ \hline
		$q(v),\bbq_{v}$,\newline $\bbq $ & $q(v)=\bbq_{v}$: the queue length of link $v$, \newline $\bbq$: the vector of queue lengths on all links $\ccalV$ \\ \hline
		$ Q(S,\bbz) $ & Q-value of state-action pair ($S$, $\bbz$) in the training formulation in Section~\ref{sec:train} \\ \hline
		$ \bbQ(S,\bbz) $ & Q vector that captures the contribution of each dimension of action $\bbz$ to the Q-value $ Q(S,\bbz) $ \\ \hline
		$r(v),\bbr_{v}$,\newline $\bbr$ &  $r(v)=\bbr_{v}$: link rate of link $v$,\newline $\bbr$: the vector of link rates for all links $\ccalV$  \\ \hline
		$ S,\bbS $ & $S=(\ccalG,\bbS)$: state in the training formulation in Section~\ref{sec:train},\;\; $\bbS$: matrix of input features on all vertices $\ccalV$ \\ \hline
		$u(v)$,\newline $u(\boldsymbol{v})$ &  $u(v)$: utility value on link $v$,\newline $u(\boldsymbol{v})$: total utility on independent set $\boldsymbol{v}$  \\ \hline
		$\bbu$,\newline $\bbu_{v}$ & $\bbu$: the vector of utility values on all links $\ccalV$,\newline $\bbu_{v}=u(v)$: utility value on link $v$  \\ \hline
		$ \mathbb{U}(a,b)$ & uniform distribution between $a$ and $b$  \\ \hline
		$v$ &  $v\in\ccalV$: ID of an arbitrary vertex  (link) on the conflict graph $\ccalG$ (network)   \\ \hline
		$\boldsymbol{v}$,\newline $\boldsymbol{v}^*$ & $\boldsymbol{v}\subseteq\ccalV$: an arbitrary independent set on graph $\ccalG$,\newline $\boldsymbol{v}^*$: the optimal solution of MWIS problem   \\ \hline
		$\hat{\boldsymbol{v}}_{\mathrm{Gr}}$,\newline $\hat{\boldsymbol{v}}_{\mathrm{GCN}}$ &  $\hat{\boldsymbol{v}}_{\mathrm{Gr}}$: solution from greedy MWIS solver, \newline  $\hat{\boldsymbol{v}}_{\mathrm{GCN}}$: solution from GCN-based MWIS solver   \\ \hline
		$\bbv$,\newline $\hat{\bbv}$ &  $\bbv\in\{0,1\}^{|\ccalV|}$: the indicator vector of set $\boldsymbol{v}$ w.r.t. $\ccalV$,\newline $ \hat{\bbv} $: the indicator vector of $ \hat{\boldsymbol{v}}_{\mathrm{GCN}} $ w.r.t. $\ccalV$  \\ \hline
		$V$ &  graph size, the number of vertices on graph $\ccalG$, $V=|\ccalV|$   \\ \hline
		$\bbw$ & vector of topology-weighted utilities, $\bbw=\bbz\odot\bbu$  \\ \hline
		$\bbx,\bbx_i$ & $\bbx$: Upright bold lower-case symbol denotes a column vector,\; $\bbx_i$: the $i$th element of vector $\bbx$. \\ \hline
		$\bbX$,\newline $\bbX_{ij} $ & $\bbX$: Upright bold upper-case symbol denotes a matrix,\newline $\bbX_{ij}$: element at row $i$ and column $j$ of matrix $\bbX$. \\ \hline
		$\bbX_{i*}$,\newline $\bbX_{*j} $ & $\bbX_{i*}$: the entire row $i$ of matrix $\bbX$,\newline $\bbX_{*j}$: the entire column $j$ of matrix $\bbX$. \\ \hline
		$\bbz,\bbZ,\bbz(\ccalG)$ & node embeddings generated by GCN (based on $\ccalG$), either as a vector $\bbz\in\reals^{|\ccalV|}$ or a matrix $\bbZ\in\reals^{|\ccalV|\times B}$  \\ \hline
	\end{tabular}
	\vspace{-0.2in}
\end{table*}

\section{Related Work}\label{sec:review}

The MWIS problem has been studied for decades. Centralized solvers based on exact, heuristic, and hybrid algorithms have been developed~\cite{linderoth1999computational,woeginger2003exact,warren2006combinatorial,puchinger2005combining,dimakis2006sufficient,verhetsel2017solving,lamm2019exactly}. 
To find the exact solution, the MWIS problem can be formulated following integer programming, maximum satisfiability, or graph coloring approaches, and solved via mixed integer programming solvers \cite{gurobi} based on branch-and-bound schemes~\cite{linderoth1999computational,woeginger2003exact,warren2006combinatorial}. 
For general graphs, exact solvers only work on medium-sized graphs of up to hundreds of vertices, since the MWIS problem is NP-hard. 
For real-world graphs, however, their structural properties can be utilized to improve the efficiency of exact solvers \cite{Kabbani07,lamm2019exactly}. 
For {networks with a tree topology}, the MWIS problem can be solved efficiently~\cite{Kabbani07}. 
Moreover, a full suite of rule-based graph reduction techniques has been developed to exploit the hierarchical structure of large real-world graphs~\cite{lamm2019exactly}, which can drastically reduce the effective size of graphs being processed in iterative frameworks and allow the exact solvers to work for some real-word graphs of up to millions of vertices.
For medium to large-sized graphs that are unsolvable by the exact solvers, heuristics based on local search can often obtain approximate solutions with high quality \cite{pullan2006phased,pullan2009optimisation,wu2012multi,benlic2013breakout}. 
In addition, quantum approximate optimization combining quantum and classical computing has been recently proposed for heuristic solvers ~\cite{Lucas14Ising,choi2020energy}. 
Nonetheless, the aforementioned centralized MWIS solvers are not suitable for link scheduling due to complexity~\cite{Kabbani07}, since centralized link scheduling generally requires solving the MWIS problem on small graphs within milliseconds. Our centralized MWIS solver is specifically designed to work on general graphs at low complexity {in terms of time and communication.}

For practical scheduling in wireless ad-hoc networks, distributed MWIS solvers with low communication and computational complexity are usually preferred.
Distributed MWIS solvers \cite{sanghavi2009message,paschalidis2015message,du2016new,Douik2018,Lucas14Ising,Li18Ising,Ameen20Ising} construct a solution through an iterative procedure of a round of local exchanges between a vertex and its neighbors, followed by a phase of processing on each vertex. 
Thus, it is best to describe the complexity of distributed MWIS solvers with \textit{local (communication) complexity}, defined as the number of rounds of local exchanges between each vertex and its neighbors.
In \cite{paschalidis2015message,sanghavi2009message,Douik2018}, a solution is obtained in 2 steps, each with a linear local complexity of $\ccalO(V)$: first, solve the linear relaxation of the integer programming formulation of the MWIS problem with clique constraints, then use the solution of step 1  as initial weights to estimate the solution in a distributed manner.
In \cite{du2016new}, a solution is constructed through $\ccalO(V)$ iterations of combining feasible local solutions at each vertex and exchanging the results with its neighbors.
Compared with distributed solvers  with linear local complexity $\ccalO(V)$ \cite{sanghavi2009message,paschalidis2015message,du2016new,Douik2018}, Ising-formulated MWIS solvers \cite{Lucas14Ising,Li18Ising,Ameen20Ising} require a fixed number of rounds (e.g. tens to hundreds) of local exchanges to emulate the cooling process of atoms with magnetic spin. 
The distributed greedy solvers \cite{joo2012local,joo2015distributed} have an average local complexity of $\ccalO(\log V)$, and the worst-case local complexity of $\ccalO(V)$ on certain graphs. 
In particular, the local greedy solver (LGS) \cite{joo2012local} selects vertices with the largest weights among their neighbors with a built-in tie resolution mechanism, and then excludes the neighbors of the selected vertices. 
Randomization is introduced in \cite{joo2015distributed} to improve the {complexity} of LGS, which is deterministic; a vertex is selected if its weight exceeds a prescribed fraction of the maximal weight of its neighbors. 
Our distributed MWIS solver departs from existing work by incorporating topological information in the solution through a trainable node embedding procedure.

GNNs have been recently proposed to approximate the solution to combinatorial problems~\cite{khalil2017learning, mittal2019learning} including the maximal (unweighted) independent set (MIS) problem~\cite{li2018combinatorial}, the Boolean satisfiability problem~\cite{selsam2018learning}, the traveling salesman problem~\cite{prates2019learning}, and the maximum constraint satisfaction problems~\cite{tonshoff2020graph}. 
These learning-based solvers often prioritize suboptimality gap over time complexity, while ours do the opposite.
{The GCN-guided tree search in~\cite{li2018combinatorial} randomly traverses a search tree predicted by a GNN, for many times, in order to find as many candidate solutions as possible, and outputs the best one at timeout.
Similar to~\cite{li2018combinatorial}, our centralized solver also builds its solution by traversing a search tree predicted by a GNN.
However, we adopt a rollout strategy \cite{bertsekas2021rollout} to further guide the tree traversal so that we can find a good solution fast by traversing the tree only once.}
Moreover, the aforementioned works \cite{khalil2017learning,mittal2019learning,li2018combinatorial,selsam2018learning,prates2019learning} propose centralized {solvers} whereas, to the best of our knowledge, we provide the first fully distributed GCN-based {solver} to the MWIS problem.
In addition, our results demonstrate meaningful contributions of GNN to the enhanced quality of our centralized and distributed solvers, whereas the contribution of GNN to the tree-search in  \cite{li2018combinatorial} is questionable \cite{bother2022s}.
Finally, we develop a customized RL scheme to train the GCN embedded in a {non-differentiable} pipeline, which complements the existing approaches of zeroth-order optimization \cite{liu2020primer} and surrogate gradient \cite{poganvcic2019differentiation}.

\section{System Model and Problem Statement}
\label{sec:problem}

\subsection{System Model}\label{sec:problem:sys}
\begin{figure}[t]
    \vspace{-0.15in}
	\centering
   	\subfloat[]{
		\includegraphics[width=0.5\linewidth]{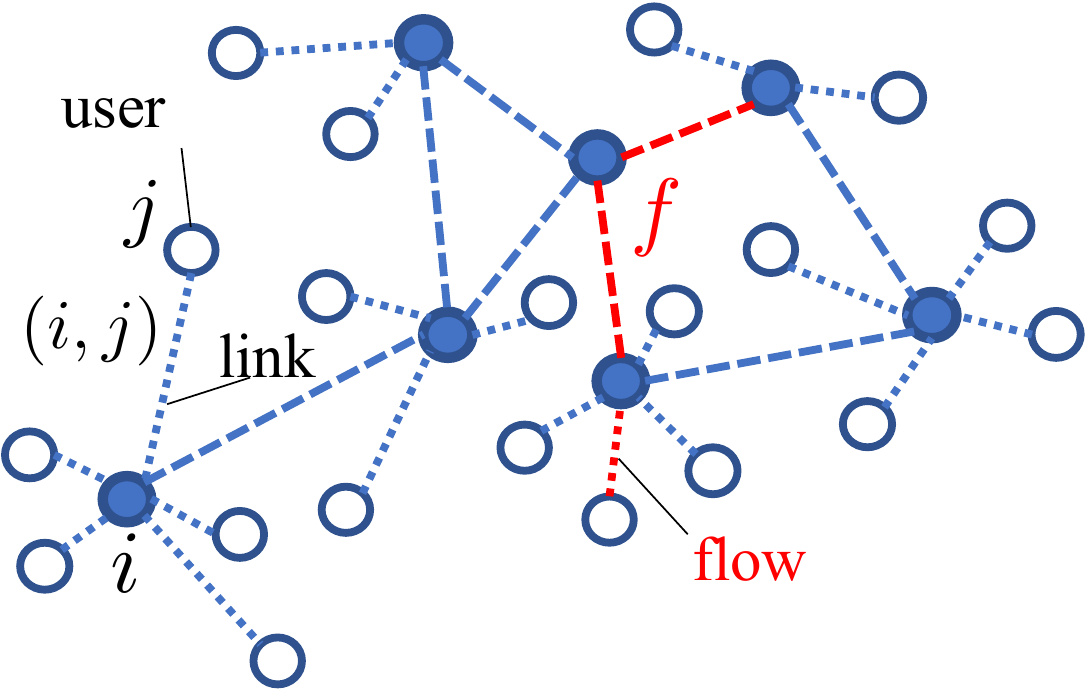}
		\label{fig:mhw}
	}%
	\subfloat[]{
		\includegraphics[width=0.4\linewidth]{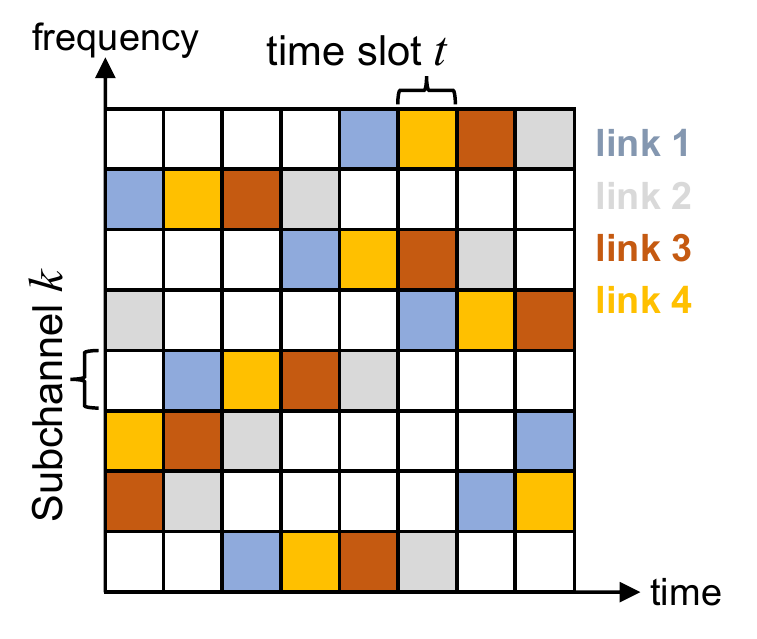}
		\label{fig:fdma}
    }  
	\vspace{-0.05in}
    \caption{Wireless multihop network with orthogonal access. (a)~Connectivity graph of the network. (b)~Example of orthogonal access in an FDMA system, where the spectrum is divided into sub-channels, time is divided into time slots, and each spectral-temporal slot can be accessed by at most one link in a set of potentially interfering links.} \label{fig:system}
	\vspace{-0.15in}
\end{figure}

Consider a wireless multihop network as illustrated in Fig.~\ref{fig:mhw}, where the existence of link $(i,j)$ implies that user $i$ and user $j$ can communicate with each other. 
Since we will ultimately focus on a conflict graph whose vertices represent links in the wireless network, we denote an arbitrary link $(i,j)$ as $v$.
A flow $f$ describes the stream of packets from a source user to a destination user. 
A flow may pass through multiple links determined by a routing scheme. 
For each link, there is a queuing system $q$ for packets of all the flows.

The wireless network adopts a multiple access scheme that divides the spectrum resource into a set of orthogonal sub-channels, $\ccalK=\{1,\dots,K\}$, and time slots.
The wireless channel is assumed to be stationary and ergodic network wide, and invariant within a time slot, i.e., the coherence time of the channel is assumed to exceed the duration of a time slot.
In Fig.~\ref{fig:fdma}, an example of a frequency division multiple access (FDMA) system is illustrated.\footnote{Alternatively, in code division multiple access (CDMA) systems, channels are implemented by orthogonal codes.} 
The channel state information of link $v$ on sub-channel $k$ at time slot $t$ is denoted by $h^{(t)}_{k}(v)$. 
We assume an orthogonal access scheme where each slot of the spectral-temporal grid in Fig.~\ref{fig:fdma} can only be accessed by one link out of a set of potentially interfering links. 

There are two formulations of interference constraints in the literature: physical distance model and hop distance model \cite{cheng2009complexity}. 
In this work, the interference in the system is considered to follow a physical distance model. 
For example, two links interfere with each other if their incident users are within a certain distance such that their simultaneous transmission will cause the outage probability to exceed a certain level.  
Depending on the air-interface technology and antenna systems, the interference zone of a link can be different from its connectivity zone. 
A link is assumed to be able to learn its interfering neighbors by monitoring the channel and/or beacon signals.
Moreover, mutually interfering links are assumed to be able to exchange control messages, e.g., with low-rate modulation and coding schemes. 
Notice that, in principle, the interference zone of a link would depend on the transmit power of the corresponding user and hence possibly vary with time.
To simplify the analysis and avoid this dependence, we consider a scenario in which all the users transmit at power levels that do not vary with time. 
In general, our approaches work on any conflict graph, no matter how it is constructed. 

\subsection{Single-Radio Single-Channel Scheduling}
We first consider a wireless network with only one sub-channel, in which each user is equipped with one half-duplex radio interface.
The interference relationship between links of the wireless multihop network is described by a \emph{conflict graph} $\ccalG$, where a vertex in the conflict graph represents a link in the wireless network and the presence of an edge in $\ccalG$ encodes the fact that the corresponding links interfere with each other.
In the rest of this paper, we focus on the conflict graph $\ccalG$ which we assume to be known; see, e.g.,~\cite{yang2016learning} for its estimation, {and the supplemental materials \cite[Sec.~V]{supplement} for the complexity of its construction.} 
Recall that an independent (vertex) set (IS) in a graph is a set of vertices such that there are no edges between any two vertices in the set.
From the definition of $\ccalG$, only wireless links that form an IS in $\ccalG$ can communicate simultaneously in time and frequency under the constraint of orthogonal access.

The state of the system at time slot $t$ can be described by the tuple $(\ccalG^{(t)}, \mathbf{q}^{(t)}, \mathbf{f}^{(t)}, \mathbf{h}^{(t)})$ consisting of the conflict graph $\ccalG^{(t)}$, queue lengths $\mathbf{q}^{(t)}$, flows $\mathbf{f}^{(t)}$, and channel states $\mathbf{h}^{(t)}$.
Since the scheduling is conducted at each time slot $t$, for notational simplicity we omit the superscript $t$, denoting the system state as $(\ccalG, \mathbf{q}, \mathbf{f}, \mathbf{h})$. 
In this setting, the task of optimal link scheduling can be described as selecting a set of non-interfering links, $\boldsymbol{v}\subseteq\ccalV$, on which to transmit in order to maximize some utility $u(\boldsymbol{v}) = f(\boldsymbol{v}; \ccalG, \mathbf{q}, \mathbf{f}, \mathbf{h})$ that is parameterized by the current state of the system.
As is customary \cite{marques2011optimal}, we model here the utility of the set of links $\boldsymbol{v}$ (or the set of vertices in the conflict graph) as the sum of utilities associated with each link, i.e., $u(\boldsymbol{v}) = \sum_{v\in\boldsymbol{v}}u(v)$, leading to the following formal problem statement.

\begin{Problem}\label{P:main}
    Consider a conflict graph $\ccalG(\ccalV,\ccalE)$, where $\ccalV$ and $\ccalE$ describe all the links and their conflict relationships in the wireless network, respectively, and a utility function $u: \ccalV \to \reals_+$. The optimal scheduling is given by selecting a subset of vertices $\boldsymbol{v}^* \subseteq \ccalV$ such that
	\begin{subequations}\label{eq:mwis}
		\begin{align}
		& \boldsymbol{v}^* = \argmax_{\boldsymbol{v}\subseteq \ccalV} \,\, \sum_{v\in\boldsymbol{v}} u(v) \\
		\text{s.t. } & (v_i, v_j)\notin \ccalE\;, \forall \, v_i, v_j \in \boldsymbol{v}. \label{eq:mwis:constraint}
		\end{align}
	\end{subequations}
\end{Problem}

With the statement of Problem~\ref{P:main}, the optimal scheduling at each temporal slot is transformed to an MWIS problem in the corresponding conflict graph.
Indeed, we want to choose non-neighboring vertices in the conflict graph (i.e., non-interfering links in the wireless network) such that the total utility is maximized. 
As discussed in Section~\ref{sec:intro} and formally introduced here, this utility $u$ is a function of the current state of the network with many existing variants~\cite{Joo09,joo2012local,Douik2018,joo2015distributed,paschalidis2015message,marques2011optimal}.
The \textit{MaxWeight} scheduler described by Problem~\ref{P:main} is not tied to a specific utility function, but rather can work with any utility function.

\begin{figure}[t!]
	\begin{minipage}[t]{\linewidth}
		\vspace{-0.2in}
		\begin{algorithm}[H]
			\setstretch{1.0}	
			\caption{{Local greedy solver $\hat{\boldsymbol{v}}_{\mathrm{Gr}} = g_d(\ccalG,\bbu)$~\cite{joo2012local}}}
			\label{algo:lgs}
			\hspace*{\algorithmicindent} \textbf{Input:} $\ccalG, \bbu$ \\
			\hspace*{\algorithmicindent} \textbf{Output:} $\hat{\boldsymbol{v}}_{\mathrm{Gr}}$ 
			\begin{algorithmic}[1] 
				\STATE $\hat{\boldsymbol{v}}_{\mathrm{Gr}}\gets\varnothing$; $\ccalG'(\ccalV',\ccalE')\gets\ccalG(\ccalV, \ccalE)$; $\bbc=\boldsymbol{0}$
				\WHILE{$\ccalG' \neq \phi$}
				\FORALL{ $v\in \ccalV'$ }
				\STATE $v$ exchanges $u(v)$ with its neighbors $\forall v_i \in \ccalN_{\ccalG'}(v)$
				\IF{ $ u(v) > \underset{v_i \in \mathcal{N}_{\ccalG'}(v)}{\max}\, u(v_i) $ } 
				\STATE $c(v) \gets +1$;\, $v$ broadcasts a control message
				\STATE $c({v_i})\gets -1, \forall v_i \in \mathcal{N}_{\ccalG'}(v)$
				\ENDIF
				\ENDFOR
				\STATE $\hat{\boldsymbol{v}}_{\mathrm{Gr}} \gets \{v | v\in\ccalV', c(v)=1\}$
				\STATE $\ccalV'\gets \{ v | v\in \ccalV', c(v)=0\}$, update $\ccalG'$ with new $\ccalV'$
				\ENDWHILE
			\end{algorithmic}
		\end{algorithm}
		\vspace{-0.3in}
	\end{minipage}
\end{figure}

\subsection{Multi-Channel Scheduling}\label{sec:problem:multi}
There are two ways to extend the MaxWeight scheduler defined in Problem~\ref{P:main} to networks with a set of orthogonal sub-channels $\ccalK$: 
1) Sequentially solve a set of single-channel scheduling tasks with state $(\ccalG^k, \mathbf{q}^k, \mathbf{f}, \mathbf{h}^k)$ for each sub-channel $k\in\ccalK$, where the queue lengths $\bbq^k$ for sub-channel $k$ depend on the schedules on sub-channels $\{1,\dots k-1\}$.
The complexity and scheduling overhead of this approach grows linearly with the number of channels $K$. 
2) Solve the MWIS problem on a single multi-channel conflict graph, for which detailed construction methods can be found in \cite{lin2007distributed,bhandari2010scheduling}. 
The multi-channel conflict graph is $K$ times larger than the single-channel conflict graph.
With a heuristic solver of logarithmic complexity, the second approach can reduce the complexity from $\ccalO(K\log V)$ to $\ccalO(\log KV)$ at the cost of poorer relative performance on larger graphs as illustrated in Section~\ref{sec:results:throughput}. 
In this paper, we focus on single-channel scheduling while numerically evaluating the second approach for multi-channel scheduling.

\subsection{Greedy Heuristics}
\label{sec:problem:greedy}
A centralized greedy solver (\emph{CGS}) \cite{dimakis2006sufficient}, denoted as $\hat{\boldsymbol{v}}_{\mathrm{Gr}} = g_c(\ccalG,\bbu)$, estimates $\hat{\boldsymbol{v}}_{\mathrm{Gr}}$ that approximates the solution to~\eqref{eq:mwis}  in an iterative fashion by first adding to $\hat{\boldsymbol{v}}_{\mathrm{Gr}}$ the vertex with the largest utility, deleting its neighbors as potential candidates, and repeating this procedure until all vertices are either added to $\hat{\boldsymbol{v}}_{\mathrm{Gr}}$ or deleted, as detailed in Algorithm~1 in \cite{supplement}. 

The distributed implementation of CGS is denominated as local greedy solver (\emph{LGS})~\cite{joo2012local}, denoted as $\hat{\boldsymbol{v}}_{\mathrm{Gr}} = g_d(\ccalG,\bbu)$. 
As detailed in Algorithm~\ref{algo:lgs}, if a vertex $v$ has the largest weight in the neighborhood (line 5), it is marked as $+1$ (line 6) and added to the solution set $\hat{\boldsymbol{v}}_{\mathrm{Gr}}$ (line 10), then $v$ broadcasts a control message to its neighbors, who then mark themselves as $-1$ (line 7). 
Next, the unmarked vertices form the residual graph $\ccalG'$ (line 12). 
In practice, the LGS has a built-in tie-breaking mechanism based on an initial assignment of identification numbers to each vertex in the conflict graph that does not require additional information exchanges in the case of a tie (in line 5). 
Notice that these local exchanges are between vertices in the conflict graph.
By construction, $\hat{\boldsymbol{v}}_{\mathrm{Gr}}$ is guaranteed to follow the IS constraint in \eqref{eq:mwis:constraint} but the suboptimality gap $u(\boldsymbol{v}^*) - u(\hat{\boldsymbol{v}}_{\mathrm{Gr}})$ might be large since CGS and LGS do not fully consider the topology of $\ccalG$. 

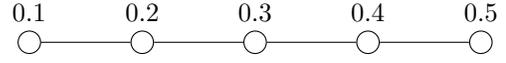
\begin{figure}[t]
	\vspace{-0.1in}
	\centering
	\begin{tikzpicture}
		\centering
		\def \n {5}
		\foreach \s in {1,...,\n}
		{
			\node[draw, circle, label=$0.\s$, inner sep=-3](a\s) at (1.5*\s,1) {};
		}
		\foreach \s/\d in {1/2,2/3,3/4,4/5}
		{
			\draw (a\s) -- (a\d);
		}
		
	\end{tikzpicture}
	\caption{An example path graph with 5 vertices of increasing weights on which LGS requires 5 iterations to complete as the worst case.}
	\label{fig:pathgraph}
	\vspace{-0.1in}
\end{figure}

Both CGS and LGS have linear computational complexity $\ccalO(V)$. 
However, as distributed MWIS solvers are parallel by nature, we focus on their local complexity. 
The LGS has a logarithmic average local complexity $\ccalO(\log V)$ on random graphs and a linear worst-case complexity $\ccalO(V)$ on path graphs with increasing vertex weights along the path as illustrated in Fig.~\ref{fig:pathgraph}~\cite{joo2012local}.

\begin{figure}[t!]
	\centering
	\includegraphics[height=2.4in]{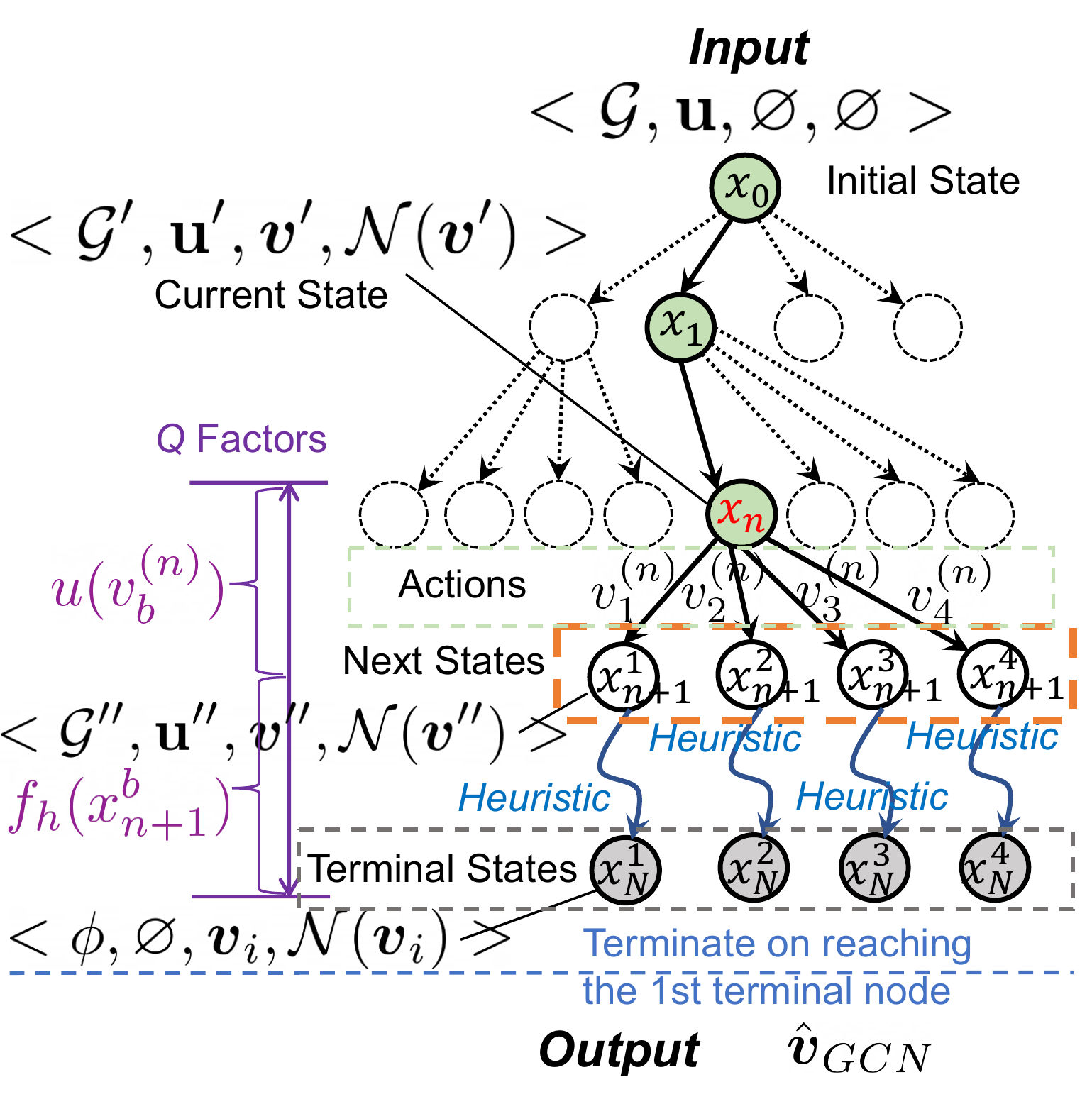}
	\caption{Exemplary search tree for the GCN-based centralized MWIS solver, where 
		the green nodes illustrate the traversal path of 1-step lookahead rollout search that can quickly reach a good terminal node. 
	}
	\label{fig:rollout}
	\vspace{-0.1in}
\end{figure}

\section{Graph convolutional network-guided tree search}
\label{sec:search}

{An enhanced centralized scheduler could improve the performance of infrastructure-based wireless multihop networks of small to medium sizes, such as D2D communications~\cite{Kim2016d2d}, CRAN \cite{Douik2018}, wireless backhaul networks~\cite{Kabbani07,gupta2020learning}, and  multihop relay networks in mmWave and THz bands~\cite{qiao2011enabling,xia2017cross}.}
Our centralized MWIS solvers are based on a unified algorithmic framework of GCN-guided tree search, 
inspired by the methodology for the unweighted MIS problem from~\cite{li2018combinatorial}. 
Our centralized solvers estimate an approximate solution $\hat{\boldsymbol{v}}_{GCN}$ by iteratively adding one vertex at a time under the guidance of a GCN without violating the constraint of independent set in \eqref{eq:mwis:constraint}. 
In the rest of this section, we first describe our GCN-guided rollout search, with a focus on the algorithmic framework, then briefly introduce two additional reference solvers following the similar framework, while leaving their details in the supplemental materials \cite{supplement}.

\subsection{Algorithmic Framework}\label{sec:search:framework}
Our methodology consists of defining a search tree of candidate solutions and then traversing it
iteratively with multiple strategies.
The search is formulated as a Markov decision process, of which the possible state transitions of each iteration form a search tree, as illustrated in Fig~\ref{fig:rollout}, which we explain in more detail in this section. 
The solver finds an approximate solution by traversing from the root of the search tree to a terminal node.
A node in the search tree represents an intermediate search state $x_{n}=<\ccalG',\bbu',\boldsymbol{v}',\ccalN_{\ccalG}(\boldsymbol{v}')>$, where $\ccalG'$ is the residual graph, $\bbu'$ is the corresponding residual utility vector collecting the utilities $u(v)$ for all $v \in \ccalG'$,  $\boldsymbol{v}'$ is the partial solution for $\ccalG$, and $\ccalN_{\ccalG}(\boldsymbol{v}')$ is the set of all the vertices adjacent to some vertex in the partial solution. 
A state transition would be triggered by the action of adding a vertex in $\ccalG'$ to the partial solution $\boldsymbol{v}'$.
The root node of the search tree,  $x_0=<\ccalG,\bbu,\varnothing,\varnothing>$, is the initial state generated from the input conflict graph $\ccalG$ and corresponding utility vector $\bbu$. 
At a terminal node, the residual graph is empty $\ccalG'=\phi$ and  $\boldsymbol{v}'$ is a feasible approximate solution to Problem~\ref{P:main}. 
Each non-terminal node has $B$ children nodes, where $B$ is the branching factor of the search tree as further discussed in Sections~\ref{sec:search:rollout} and~\ref{sec:search:random}. 
Next, we explain the iterative traversal of the search tree through the exemplary rollout strategy in Section~\ref{sec:search:rollout}.

\begin{figure}[t]
	\begin{minipage}[t]{\linewidth}
	\vspace{-0.1in}

\begin{algorithm}[H]
\caption{{GCN-guided Centralized Rollout Search}}
\label{algo:gcn-crs}
\hspace*{\algorithmicindent} \textbf{Input}: $\ccalG, \bbu$ \\
\hspace*{\algorithmicindent} \textbf{Output}: $\hat{\boldsymbol{v}}_{\mathrm{GCN}}$ 
\begin{algorithmic}[1] 
\STATE Initialize search queue $\ccalQ \gets \{x_0=<\ccalG,\bbu,\varnothing,\varnothing>\}$
\WHILE{ $\ccalQ \neq \varnothing $ }
\STATE $x_{n}=<\ccalG',\bbu',\boldsymbol{v}',\ccalN_{\ccalG}(\boldsymbol{v}')> = \mathrm{Pop}(\ccalQ)$ \COMMENTS{S.1}
\STATE $\bbz' = \Psi_{\ccalG'}(\bbu'; \bbXi)$ \COMMENTS{S.2: node embedding}
\STATE $\bbw'=\bbz'\odot\bbu'$ \COMMENTS{S.3: topology-aware utility}
\STATE $\tilde{\bbv}' = \mathrm{ArgSortDescending}(\bbw')$ \COMMENTS{S.4: argsort}
\FORALL{$ b \in \{1,\dots, B\} $ }
\STATE ${v_b^{(n)}} = \tilde{\bbv}'_b$ \COMMENTS{S.5: branching}
\STATE $\ccalG''_{b}=\ccalG'\backslash ({\{v_b^{(n)}\}}\cup\ccalN_{\ccalG'}({v_b^{(n)}})),\;\boldsymbol{v}''_{b}=\boldsymbol{v}'\cup\{{v_b^{(n)}}\}$
\STATE $\bbu''_b = [u(v)|v\in\ccalG''_b] $ 
\STATE $ x_{n+1}^{b} = <\ccalG''_{b},\bbu''_{b},\boldsymbol{v}''_{b},\ccalN_{\ccalG}(\boldsymbol{v}''_{b})> $
\STATE $ Q(x_{n+1}^{b}) = u({v_b^{(n)}}) + f_h (x_{n+1}^{b}) $ \COMMENTS{S.6: rollout}
\ENDFOR 
\STATE $<\ccalG'',\bbu'',\boldsymbol{v}'',\ccalN_{\ccalG}(\boldsymbol{v}'')> = \underset{x\in\{x_{n+1}^{b} | b \in \left[1,\dots, B\right]\}}{\argmax}\,{Q(x)}$
\IF{ $ \ccalG''\neq \phi $ }
\STATE $\ccalQ\gets\ccalQ\cup\{<\ccalG'',\bbu'',\boldsymbol{v}'',\ccalN_{\ccalG}(\boldsymbol{v}'')>\}$ \COMMENTS{S.7}
\ELSE 
\STATE $\hat{\boldsymbol{v}}_{\mathrm{GCN}}\gets \boldsymbol{v}''$
\ENDIF
\ENDWHILE
\end{algorithmic}
\end{algorithm}
\vspace{-0.2in}
\end{minipage}
\end{figure}

\subsection{GCN-guided Centralized Rollout Search}\label{sec:search:rollout}

In order to reach a good solution, the branching factor $B$ should be configured to create a large search tree with a large search space.
In the approach outlined so far, it may take a long time for the solver to find a good terminal node in a large search space.
To reach a good terminal node quickly, we introduce GCN-guided centralized rollout search (\emph{GCN-CRS}), which employs 1-step rollout search \cite{bertsekas2021rollout} to further guide the tree search, as illustrated in Fig.~\ref{fig:rollout} and Algorithm~\ref{algo:gcn-crs}.

Algorithm~\ref{algo:gcn-crs} is explained as follows.
On initialization (line 1), the root node $x_0=<\ccalG,\bbu,\varnothing,\varnothing>$ is pushed into the search queue $\ccalQ$.
In each iteration, the solver predicts $B$ vertices based on the current state (steps $1$-$5$), estimates the Q-value of adding each predicted vertex to the partial solution (step 6), and then proceeds with the action of the highest Q-value (step 7).
In step~1, a non-terminal node  $x_n=<\ccalG',\bbu',\boldsymbol{v}',\ccalN_{\ccalG}(\boldsymbol{v}')>$ is randomly popped from the search queue as the current state.
In GCN-CRS, $x_n$ is the \emph{only} item in the search queue, but other algorithm variations discussed in Section~\ref{sec:search:random} can have more items in the queue.
In step~2, the residual graph $\ccalG'$ and its corresponding utility vector $\bbu' = [u(v)|v\in\ccalG']$ are mapped by an $L$-layered GCN to a node embedding vector, as $\bbz' = \Psi_{\ccalG'}(\bbu'; \bbXi)$, where $\bbz'=\left[z'(v)|v \in \ccalV'\right] \in \reals^{V'}$ contains the topology-aware scaling factors of all vertices in $\ccalG'$, $\Psi_{\ccalG'}$ is the GCN defined on the graph $\ccalG'$ (as detailed in Section~\ref{sec:search:gcn}), and $\bbXi$ represents the collection of trainable parameters of the GCN.
In step~3 (line 5 of Algorithm~\ref{algo:gcn-crs}), a vector of topology-aware utilities of all vertices in $\ccalG'$ is created through the element-wise product $\bbw'=\bbz'\odot\bbu'$.
In step 4 (line 6), vector $\bbw'$ is sorted in descending order into $\tilde{\bbw}'$, and vector $\tilde{\bbv}'$ collects the vertices corresponding to the sorted vector $\tilde{\bbw}'$.
In step 5 (lines 7-11), the first $B$ vertices in $\tilde{\bbv}'$ (with the largest topology-aware utilities) are used to predict $B$ respective children nodes (next states) of the node $x_n$ (current state).
A child node $<\ccalG''_{b},\bbu''_{b},\boldsymbol{v}''_{b},\ccalN_{\ccalG}(\boldsymbol{v}''_{b})>$ is predicted by vertex $v_{b}^{(n)}=\tilde{\bbv}'_{b}, \forall b\in\{1,\dots,B\}$ as:
\begin{equation}\label{eq:child}
\ccalG''_{b}\!=\!\ccalG'\backslash (\{v_{b}^{(n)}\}\cup\ccalN_{\ccalG'}(v_{b}^{(n)})),
\boldsymbol{v}''_{b}\!=\!\boldsymbol{v}'\cup\{v_{b}^{(n)}\}\;.
\end{equation}
In step 6 (line 12), the Q-value of the $b$th child node $x^b_{n+1}$ (the action of adding $v_{b}^{(n)}$ to the partial solution) is estimated by 1-step lookahead rollout as $ Q(x_{n+1}^{b}) = u(v_b^{(n)}) + f_h (x_{n+1}^{b}) $.
Here, $ u(v_b^{(n)}) $, the utility of vertex $v_b^{(n)}$, is the immediate reward of transitioning from the node $x_n$ to node $x^b_{n+1}$.
The score of the $b$th child node $f_h(x^b_{n+1})$ is the total utility of a solution to the MWIS problem defined on the residual graph ($\ccalG''_{b}, \bbu''_{b}$), obtained by an efficient \emph{guiding heuristic}, which serves as the estimated reward of traversing from state $x^b_{n+1}$ to a terminal state.
Lastly, in step 7 (line 14-19), the GCN-CRS proceeds to the child node with the largest Q-value, and ties are broken randomly. 
If the selected child node (next state) is a non-terminal node, it is pushed into the search queue.
Otherwise, the solver outputs $\boldsymbol{v}''$ as the full solution.

In GCN-CRS, the search tree is traversed along a single path and the search terminates when it reaches the first terminal node.
The performance of the GCN-CRS is guaranteed to be no worse than (and largely dictated by) the guiding heuristic \cite{bertsekas2021rollout}. 
In our case, the CGS is selected as the guiding heuristic due to its linear complexity and determinism. 
Specifically, the score of the $b$th child node is obtained as $f_h(x^b_{n+1})=u({\boldsymbol{v}}^{b}_\mathrm{Gr})$, where ${\boldsymbol{v}}^{b}_\mathrm{Gr}$ is obtained by either the vanilla CGS as ${\boldsymbol{v}}^{b}_\mathrm{Gr}=g_c(\ccalG''_b,\bbu''_b)$,
or an enhanced CGS as ${\boldsymbol{v}}^{b}_\mathrm{Gr}=g_c(\ccalG''_b,\bbw''_b)$, where $ \bbw''_b = [w'(v)|v\in\ccalG''_{b}] $, $\bbw'$ and $\ccalG''_{b}$ are defined in lines 5 and 9, respectively.
Moreover, the operations on the $B$ branches in lines 7-13 can be parallelized.
{The GCN-CRS solvers using vanilla CGS and enhanced CGS as guiding heuristics are denoted as GCN-CRS-v and GCN-CRS-e, respectively.}
The training method of the GCN is detailed in Section~\ref{sec:train}.

\subsection{Graph Convolutional Network Design}\label{sec:search:gcn} 

Our GCN has an $L$-layer structure as follows: 
Given the input feature as $\bbX^0 = \bbu'$, then $\bbz' = \Psi_{\ccalG'}(\bbu'; \bbXi) = \bbX^L$, where an intermediate $l$th layer of the GCN is given by
\begin{equation}\label{E:gcn}
	\mathbf{X}^{l} = \sigma_l \left(\mathbf{X}^{l-1}{\bbTheta}_{0}^{l}+\bbcalL \mathbf{X}^{l-1}{\bbTheta}_{1}^{l}\right).
\end{equation}
In~\eqref{E:gcn}, $\bbcalL$ is the normalized Laplacian of $\ccalG$, ${\bbTheta}_{0}^{l}, {\bbTheta}_{1}^{l} \in \mathbb{R}^{g_{l-1} \times g_{l}}$ are the trainable parameters of the $l$th layer in the collection of $\bbXi$, $g_{l-1}$ and $g_{l}$ are the dimensions of the output features of layers $l-1$ and $l$, respectively, and $\sigma_l(.)$ is the activation function. 
The dimension of the input feature is $g_{0}=1$.
The activation functions of the input and hidden layers are selected as leaky ReLUs.
By setting the dimension of the output layer as $g_L=1$ with linear activation, the GCN  generates a node embedding vector, $\bbz' \in\reals^{V'}$, as a topological scaling factor.
As explained earlier, the prediction vector is then computed as $\bbw'=\bbz'\odot\bbu'$. 
The GCN is trained by RL with the help of an efficient CGS. 
Since both our proposed centralized and distributed solutions share the same training mechanism, we defer its explanation to Section~\ref{sec:train}.
{Notice that the implementation of GCN in \eqref{E:gcn} is just one viable option, other implementations of GNNs, e.g., in \cite{wu2020comprehensive}, could also be used in our framework.}

\subsection{Reference GCN-guided Centralized Solvers} \label{sec:search:random}

Next, we propose two reference solvers with which to compare GCN-CRS: GCN-guided centralized random tree search (\emph{GCN-CRTS}) and GCN-guided centralized greedy search (\emph{GCN-CGS}). 
They are obtained by modifying the state-of-the-art solvers for the unweighted MIS problem in \cite{li2018combinatorial} and \cite{khalil2017learning}, respectively, making them compatible with our MWIS setting.
GCN-CRTS and GCN-CGS follow an iterative procedure similar to that of GCN-CRS, while employing different traversal strategies.

The idea of GCN-CRTS is to reach as many random terminal nodes as possible in a given time interval to increase its chance of finding a good solution from them.
Randomized search ensures the equal chance of reaching each terminal node, 
which is implemented by unfolding a non-terminal node randomly popped from the search queue and randomly pushing non-terminal nodes along the path to the search queue.
The GCN-CRTS is almost identical to the MIS solver in \cite{li2018combinatorial}, in which a node embedding matrix $\bbZ' \in \left[0,1\right]^{V'\times B}$ is generated by a GCN with a slightly different output structure,
with only two modifications:
1) Using $\bbW' = \bbZ' \odot \bbu' \mathbf{1}^\top$ instead of $\bbZ'$ as the prediction matrix. 
2) For computational efficiency, synthetic random graphs are used as the training data for supervised learning, where the label vectors are generated by heuristics instead of exactly solving the NP-hard problem.
Specifically, the labels are generated by selecting the best solution from two guiding heuristics: linear programming \cite{KaKo2009} and centralized greedy solver.
The implementation of the GCN-CRTS is detailed in Section~II and Algorithm~2 of~\cite{supplement}.

The GCN-CGS is modified from the deep Q network in \cite{khalil2017learning} by replacing the Node2Vec module with a GCN.
For a non-terminal node $<\ccalG',\bbu',\boldsymbol{v}',\ccalN_{\ccalG}(\boldsymbol{v}')>$ in the search tree, the GCN  takes $(\ccalG',\bbu')$ as input and generates node embedding $\bbz'\in\reals^{V'}$ as the Q-values of the action space $\ccalA'=\ccalV'$, whereas the branching factor $B=|\ccalV'|$ is no longer a hyperparameter.
The action is selected by an $\epsilon$-greedy method. 
The next state $(\ccalG'',\bbu'')$ is generated according to \eqref{eq:child}.
The search terminates upon reaching a terminal node.
More details of the implementation of GCN-CGS can be found in Section~III and Algorithm~3 of~\cite{supplement}.

\begin{table}[t!]
	\renewcommand{\arraystretch}{1.1}
	\caption{Computational complexity of centralized MWIS solvers 
	} 
	\label{tab:complexity:central}
	\centering
	\footnotesize
	\begin{tabular}{l|l}
		Algorithm  & Complexity   \\ \hline
		CGS & $ \ccalO(V) $ \\ \hline
		GCN-CRTS &  $\ccalO(LV^2g^2B^{V})$ \\ \hline
		GCN-CGS &  $\ccalO(LV^2g^2)$ \\ \hline
		GCN-CRS &  $\ccalO(LV^2g^2 + {BV^2}/{\bar{d}})$  \\ \hline
	\end{tabular}
 \vspace{-0.1in}
\end{table}

Both GCN-CRTS and GCN-CGS benefit from the flexible input dimensions of GCN and the iterative algorithmic framework, so that they can generalize to large graphs and outperform the training heuristics \cite{li2018combinatorial}.
The GCN-CGS can reach a good terminal node quickly.
However, the GCN-CRTS will take a relatively long time to find a good solution (e.g., several minutes), making it unsuitable for link scheduling.

\subsection{Computational Complexity}
The computational complexity of the $l$th layer of the GCN is $\ccalO(\bar{d}Vg_lg_{l-1})$, where $\bar{d}$ is the average degree of a vertex in the input graph.
By assuming $g_l=g, \forall l\in\{1,\dots,L\}$, an $L$-layered GCN has a computational complexity of $\ccalO(LVg^2\bar{d})$. 
In tree search, reaching a terminal node requires an average of $V/\bar{d}$ steps, therefore the computational complexity of finding a solution is $\ccalO(LV^2g^2)$. 
Without timeout, the GCN-CRTS requires an exponential complexity of $\ccalO(LV^2g^2B^{V})$ to reach all $B^{V}$ terminal nodes.
To find a good solution on graphs of hundreds of vertices, a timeout of several minutes is usually required for GCN-CRTS. 
The computational complexity of GCN-CRS is $\ccalO(LV^2g^2 + BV^2/\bar{d})$, since the guiding heuristic of CGS will be executed $B$ times on each node of the search tree, adding a complexity of $\ccalO(BV^2/\bar{d})$ with an average of $V/\bar{d}$ passes. 
The computational complexities of the presented centralized MWIS solvers are summarized in Table~\ref{tab:complexity:central}. 

\begin{figure*}[t]
	\centering
	\vspace{-0.1in}
	\includegraphics[width=0.73\textwidth]{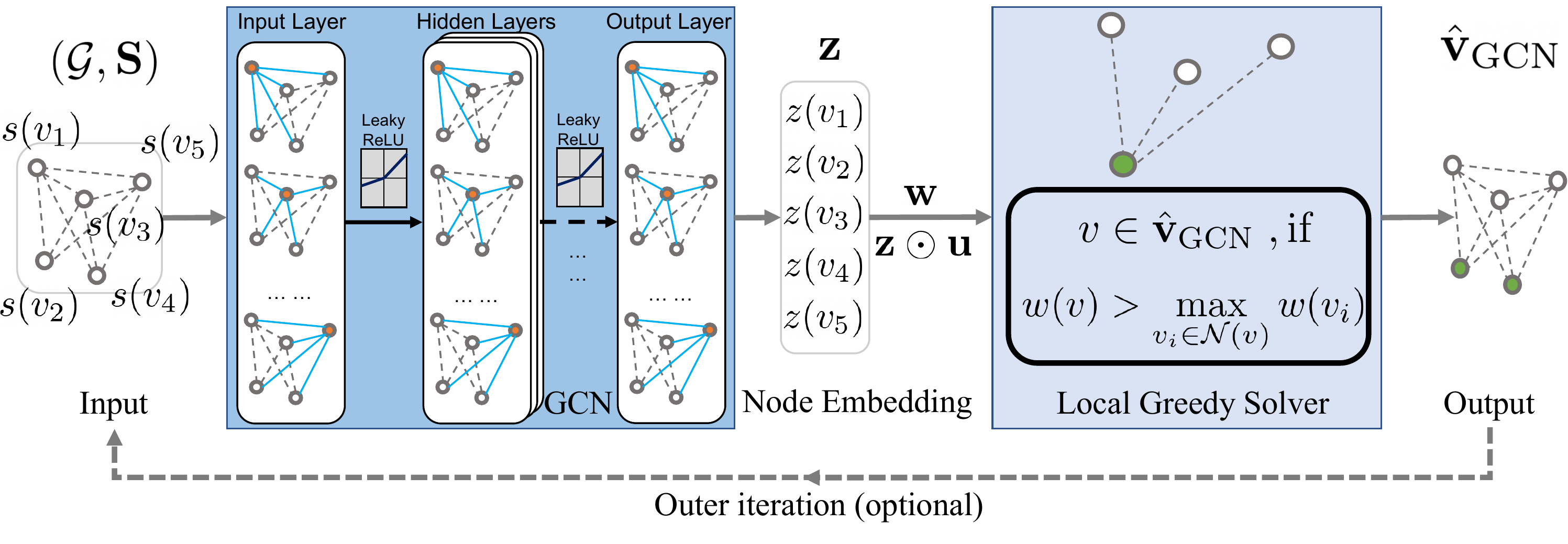}
	\vspace{-0.1in}
	\caption{Architecture of the GCN-based distributed MWIS solver. First, the conflict graph $\ccalG$ and node features $\bbS$ are encoded into the scalar embeddings $\bbz$ via a GCN. Then, the element-wise product of $\mathbf{z}$ and $\mathbf{u}$ is fed into a local greedy algorithm to generate a solution $\hat{\boldsymbol{v}}_{\mathrm{GCN}}$. 
		{The optional outer iteration represents an architectural variation, as detailed in \cite[Algo.~4]{supplement}, in which the residual graph is encoded by the GCN for each inner iteration of LGS.}
	}
	\label{fig:dqnfw}\label{fig:dqnfw:main}
	\vspace{-0.2in}
\end{figure*}

\section{Distributed MWIS solver using graph neural networks}
\label{sec:solution}

Our goal in the design of a distributed solver is to \emph{decrease the suboptimality gap of the baseline LGS described in Section~\ref{sec:problem:greedy} while keeping its two main advantages}: 1)~Low computational complexity, and 2)~Can be implemented in a distributed manner with low communication cost. 
To achieve this goal, LGS with modified weights is proposed to solve Problem~\ref{P:main}.
More precisely, mimicking the development of our centralized solvers, instead of considering the vanilla utilities $u(v)$ we consider graph-aware utilities $w(v)=z(v) u(v)$, where the scalar node embedding $z(v)$ encodes a relevant topological feature of vertex $v$.
Intuitively, if vertex $v$ is a high-degree link that may interfere with many other links in the wireless network, then $z(v)$ should downscale the utility $u(v)$ since scheduling $v$ would preclude many other links from the schedule. 
By contrast, if vertex $v$ has low-degree in the conflict graph $\ccalG$ (e.g., an isolated link in the original wireless network) then $z(v)$ should amplify $u(v)$. 
In summary, $z(v)$ should be a topology-aware scaling that reduces the MWIS suboptimality gap and, to be consistent with our goal, should also be attainable in a distributed manner with low communication and computational cost.
With these requirements in mind, we propose to obtain a vectorized node embedding $\bbz\in\mathbb{R}^{V}$ collecting $z(v)$ for all $v \in \ccalV$ as the output of a GCN~\cite{kipf2016semi}. 
The distributed MWIS solver composed of a distributed GCN followed by LGS is denoted by \emph{GCN-LGS}.
The architecture of the GCN-LGS solver is illustrated in Fig.~\ref{fig:dqnfw}. 
As detailed later, the training of GCN-LGS is centralized while its execution is distributed.

\subsection{Distributed Architecture}
\label{sec:solution:forward}

In accordance with Problem~\ref{P:main}, the inputs to the GCN-LGS solver consist of the conflict graph $\ccalG$ and a feature matrix $\bbS$, and the output of the solver is an estimate $\hat{\boldsymbol{v}}_{\mathrm{GCN}}$ of the optimal link scheduling. 
The feature matrix $\bbS$ can be the utility vector $\bbu$, a constant vector $\mathbf{1}$ (featureless case), or contain other vertex features.

The first step is to obtain a topology-aware node embedding $\bbz\in\mathbb{R}^{V}$, as $\bbz = \Psi_{\ccalG}(\bbS; \bbXi)$, where $\Psi_{\ccalG}$ is an $L$-layered GCN {defined on $\ccalG$}, 
which is the same distributable GCN used in our centralized solver as described in \eqref{E:gcn}.
Since the normalized Laplacian $\bbcalL$ is a local operator on $\ccalG$, it should be noted that $z(v)$ can be computed locally at each vertex $v$ through $L$ rounds of local exchanges with its neighbors. 
Specifically, by avoiding the use of global operations, such as network-wide softmax activations or normalizations, the system level update of the $l$th layer of a GCN in \eqref{E:gcn} can be implemented in a fully distributed manner via the following local operation on link $v\in\ccalV$,
\begin{equation}\label{eq:1layer}
    \bbX_{v*}^{l} \!=\! \sigma_l \! \left(\! \bbX_{v*}^{l-1} \, \bbTheta_{0}^{l} \!+\! \left[\! \bbX_{v*}^{l-1} \!-\!\!\! \sum_{u \in \mathcal{N}(v)}\! \frac{\bbX_{u*}^{l-1}}{\sqrt{d({v})d({u})}}\! \right]\! \bbTheta_{1}^{l} \! \right),
\end{equation}
where $\bbX_{v*}^{l}\in\reals^{1\times g_{l}}$ is the $v$th row of matrix $\bbX^{l}$ in \eqref{E:gcn}, capturing the features on vertex $v$, $\mathcal{N}(v)$ denotes the neighbor set of vertex $v$, and $d(\cdot)$ is the degree of a vertex.
{In practice, a link $v$ can track its degree by counting its interfering neighbors during local exchanges.}
The expression in~\eqref{eq:1layer} shows that the local computational complexity of link $v$ scales linearly with its degree, which is small in many practical scenarios.
The major complexity of GCN-LGS comes from local exchanges.
Next, the approximate solution is estimated as $\hat{\boldsymbol{v}}_{\mathrm{GCN}} = g_d(\ccalG,\bbw)$  through the LGS~\cite{joo2012local} detailed in Algorithm~\ref{algo:lgs}, where the graph-aware utilities $\bbw = \bbz \odot \bbu$.

\begin{figure}[!t]
    \centering
    \vspace{-0.2in}
    \includegraphics[height=1.7in]{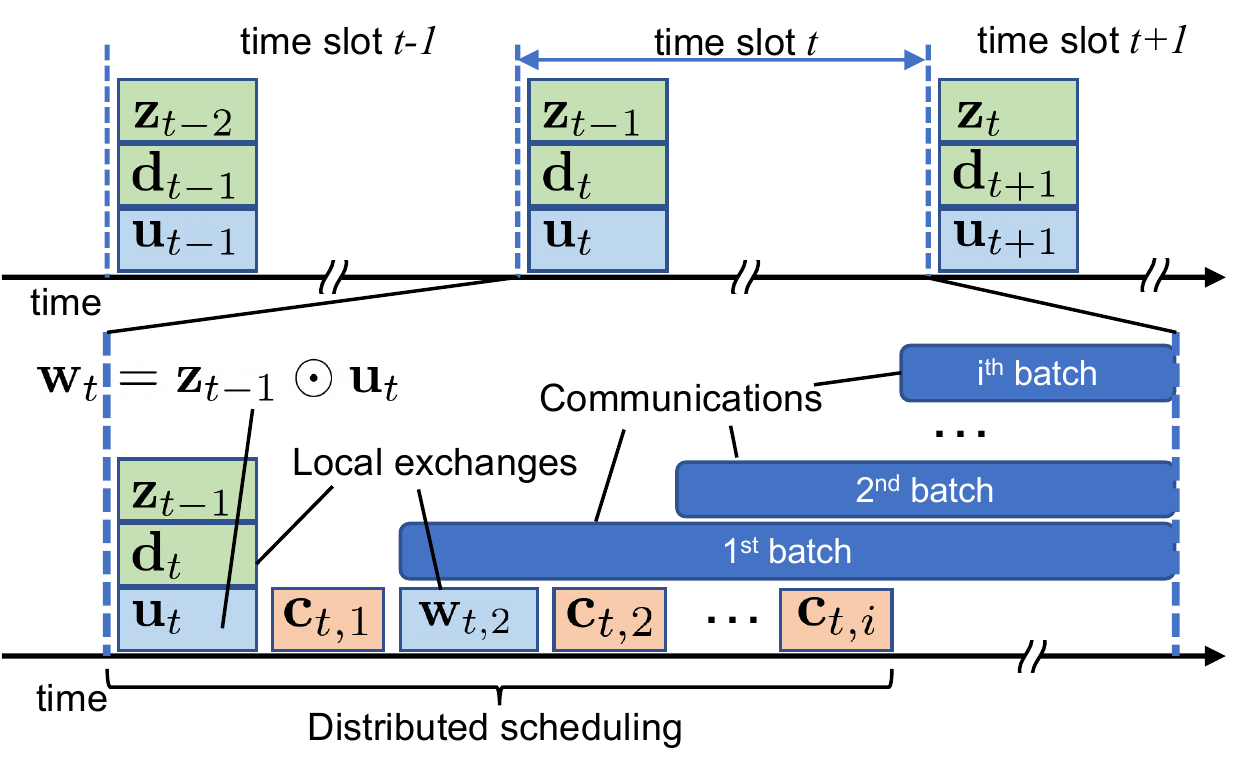}
    \vspace{-0.1in}
    \caption{Timeline of distributed scheduling with messages of 1-layer GCN piggybacked on the local exchanges of the LGS. By reusing $\bbz(\ccalG)^{(t-1)}$, the GCN-LGS has the same local complexity of LGS. 
    	In this process, node embedding $\bbz$, node degrees $\bbd$, per-link utilities $\bbu$, control messages $\bbc$, and topology-aware utilities $\bbw$ are exchanged.}
    \label{fig:timeline}
    \vspace{-0.1in}
\end{figure}

\subsection{Architectural Variations}
\label{sec:solution:reuse}

Apart from the aforementioned baseline architecture, the GCN-LGS admits several useful variations. 
The first set of variations is associated with the iterative structure, as illustrated in Fig.~\ref{fig:dqnfw:main}. 
First, the baseline LGS can be truncated to $N$ iterations, denoted as \emph{GCN-LGS-$N$}, in order to achieve a constant local complexity at the cost of the quality of solution, 
which is a key aspect to promote scalability since the constant local complexity is independent of the graph size $V$. 
In the second variation, denoted as \emph{GCN-LGS-it}, the GCN can be placed before each inner iteration of LGS ({i.e., before line 3 in  Algorithm~\ref{algo:lgs}}), so that the input of the LGS iteration $\bbw'=\bbz'\odot\bbu'$, where $\bbz'=\Psi_{\ccalG'}(\bbu';\bbXi)$, is based on the residual graph $(\ccalG',\bbS')$ from the previous iteration, rather than the input graph $(\ccalG,\bbS)$ as in the baseline GCN-LGS, {as detailed in Algorithm~4 in \cite{supplement}}.
The GCN-LGS-it can further improve the performance over the baseline GCN-LGS at the cost of higher local complexity.

The second variation refers to the choice of input features as $\bbS=\mathbf{1}$. 
We can think of the node embedding $\bbz=\Psi_{\ccalG}(\mathbf{1};\bbXi)$ generated by a featureless GCN as a \textit{topological embedding}, denoted as $\bbz(\ccalG)$, which can be reused until the network topology changes. 
In practice, $\bbz(\ccalG)$ can be reused for a coherent window of $T$ time slots that matches the pace of topological change to further reduce the computational and communication complexities of a distributed scheduler. 

Moreover, by reusing the topological embedding generated $L$ time slots earlier, i.e., use $\bbw^{(t)}=\bbz(\ccalG)^{(t-L)}\odot\bbu^{(t)}$ in time slot $t$, the additional local complexity of GCN for GCN-LGS can be reduced to zero, since $\bbz(\ccalG)^{(t-L)}$ and the intermediate features $\bbX^{l}$ can be piggybacked to the local exchange of $\bbd^{(t)}$ and $\bbu^{(t)}$ at the beginning of time slot $t$. 
An exemplary GCN-LGS with 1-layer GCN and $\bbw^{(t)}=\bbz(\ccalG)^{(t-1)}\odot\bbu^{(t)}$ is illustrated in Fig.~\ref{fig:timeline}. 
Compared to LGS, this GCN-LGS only incurs larger control messages for the first round of local exchange, additional local computational complexity, and slight topological mismatch between consecutive time slots.
The robustness of GCN-LGS to topological mismatch is evaluated in Section~\ref{sec:results:random}.

Note that a scheduled link $v$ in LGS can start to transmit right after broadcasting a control message $\bbc^{(t)}_{v}$ to mute its interfering neighbors, without the need to wait until every link in the network has been determined or until a maximum number of local exchanges has been reached.
Under this scheme, the impact of the scheduling overhead of LGS (and GCN-LGS) on spectrum utilization efficiency is further reduced without truncation.

\begin{table}[t!]
	\renewcommand{\arraystretch}{1.1}
	\caption{Local complexity of distributed MWIS solvers} 
	\label{tab:complexity:local}
	\centering
	\footnotesize
	\begin{tabular}{l|l|l}
		Algorithm  & Worst & Average   \\ \hline
		Local greedy solver (LGS) \cite{joo2012local} & $\ccalO(V)$ & $\ccalO(\log V)$ \\ \hline
		Threshold local greedy \cite{joo2015distributed} & $\ccalO(\log_{\alpha}(\beta V))$ & $\ccalO(\log_{\alpha}(\beta V))$ \\ \hline
		Message passing \cite{paschalidis2015message} & $\ccalO(V)$ & $\ccalO(V)$ \\ \hline
		Ising \cite{Li18Ising} & const. $\sim10^2$ & const. $\sim10^2$ \\ \hline
		GCN-LGS & $\ccalO(L+V)$ & $\ccalO(L+\log V)$ \\ \hline
		GCN(reuse)-LGS & $\ccalO(V)$ & $\ccalO(\log V)$ \\ \hline
		GCN-LGS-$N$ & $\ccalO(L+N)$ & $\ccalO(L+N)$ \\ \hline
		GCN-LGS-it & $\ccalO((L+1)V)$ & $\ccalO((L+1)\log V)$ \\ \hline
	\end{tabular}
	\vspace{-0.2in}
\end{table}

\vspace{-0.15in}
\subsection{Local Communication Complexity} 
\label{sec:solution:complexity}
The worst and average local complexities of the baseline and proposed distributed MWIS solvers are listed in Table~\ref{tab:complexity:local}. 
Without reusing the topological embedding, the average local complexity of the GCN-LGS solver is $\ccalO(L+\log V)$, where $\ccalO(L)$ is the local complexity of the GCN and $\ccalO(\log V)$ is the  average local complexity of LGS (as discussed in Section~\ref{sec:problem:greedy}). 
For LGS truncated to $N$ iterations, the local complexity of GCN-LGS-$N$ is $\ccalO(L+N)$. 
In this way, the local computational and communication costs of our distributed schedulers can be controlled by modifying the number of layers $L$ in the GCN. 
Moreover, by reusing the topological embedding, the average local complexity of GCN-LGS is further reduced to $\ccalO(\log V)$. 

\vspace{-0.1in}
\section{Centralized Training}
\label{sec:train}\label{sec:train:scheme}
Having discussed the rationale and the mechanics of the downstream architectures, we are left to discuss how to train the parameters $\bbXi$ in the GCN.
Compared to the typical supervised or semi-supervised settings in which GCNs are employed, our proposed downstream pipelines face two  challenges.
First, it is generally infeasible to obtain the optimal solution $\boldsymbol{v}^*$ (the labels for supervised learning) of a simulated training instance ($\ccalG$, $\bbS$, $\bbu$), since this would require solving an NP-hard problem.
Second, the output $\bbz$ of the GCN is related to the objective to be maximized $u(\hat{\boldsymbol{v}}_{\mathrm{GCN}})$ through a non-differentiable discrete function, e.g. CGS and LGS, which prevents the gradients w.r.t. the objective being back-propagated to the GCN.
To overcome these two challenges, we develop an RL scheme of graph-based deterministic policy gradient that trains the GCN based on the performance of our algorithm relative to an efficient greedy algorithm.

\begin{figure*}[!t]
	\centering
	\vspace{-0.2in}
	\subfloat[]{
		\includegraphics[width=0.47\linewidth]{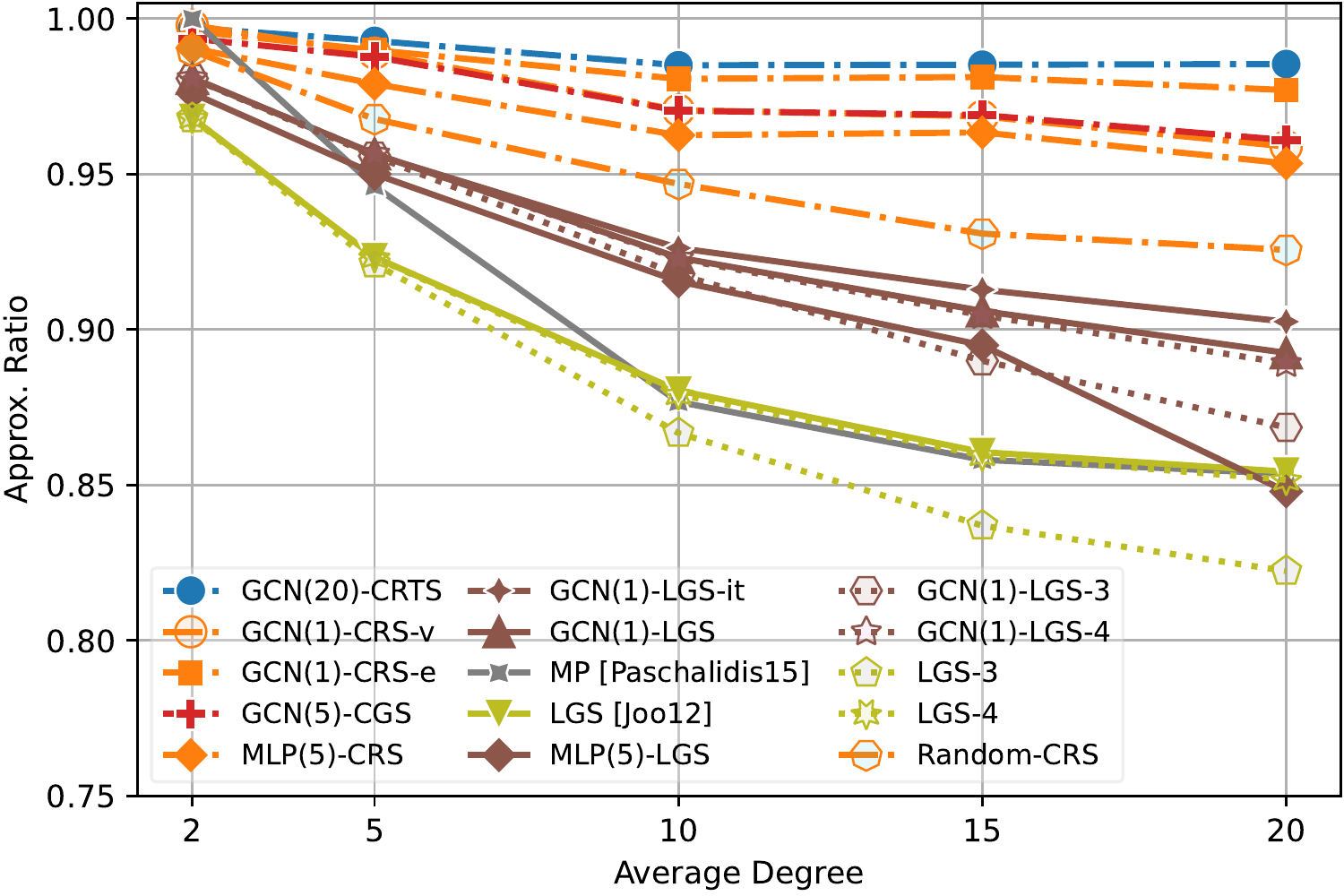}
		\label{fig:cmpx:degree}\vspace{-0.1in}
	}
	\subfloat[]{
		\includegraphics[width=0.47\linewidth]{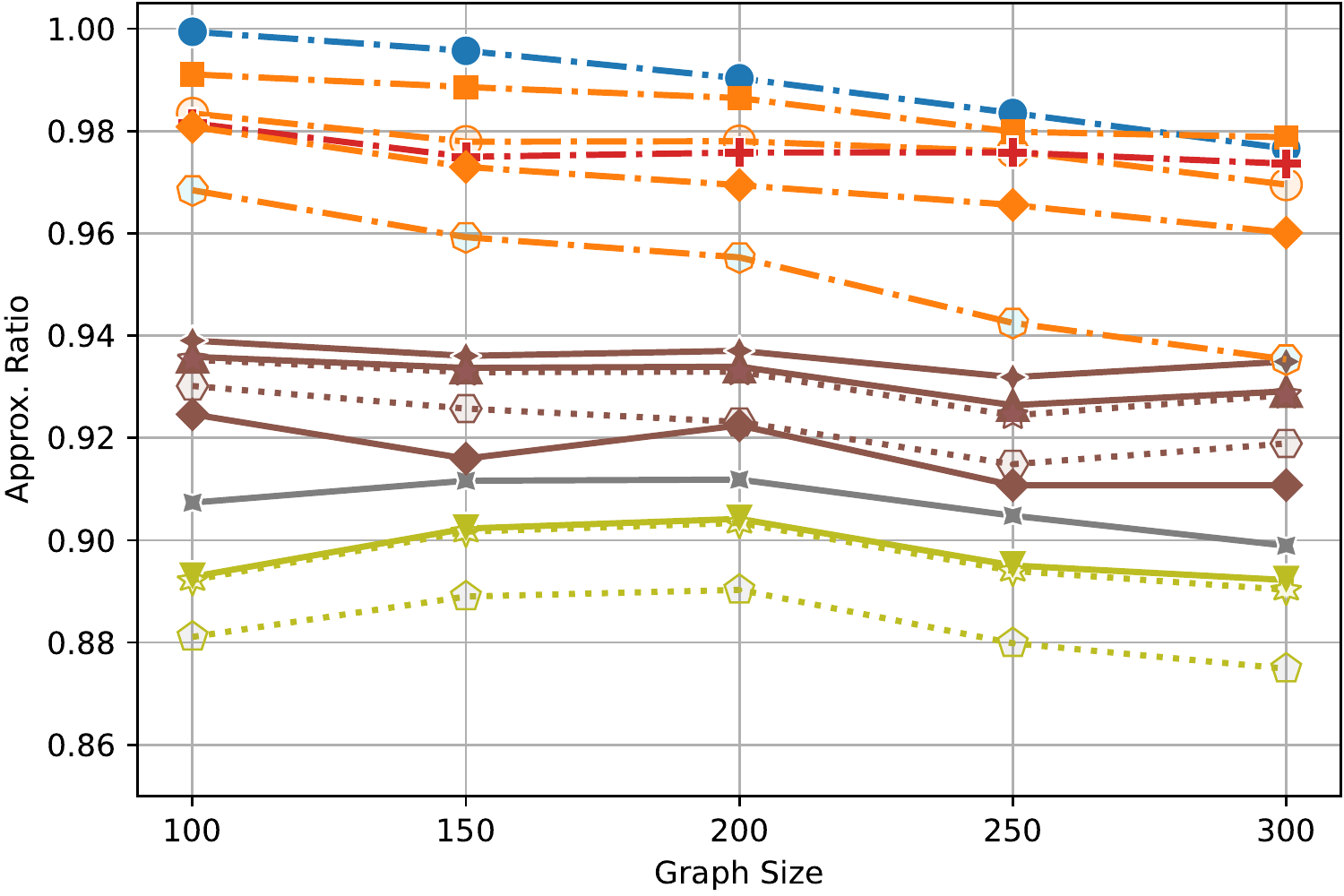}
		\label{fig:cmpx:size}\vspace{-0.1in}
	}
	\vspace{-0.05in}
	\caption{Approximation ratios of MWIS solvers on ER graphs as a function of (a)~average degree and (b)~graph size.}\label{fig:cmpx}
	\vspace{-0.15in}
\end{figure*}

We formulate the scheduling in each time slot as a \emph{single step} episode, of which the state is $S=(\ccalG,\bbS)$, the multidimensional continuous action is  $\bbz=\Psi_{\ccalG}(\bbS;\bbXi)\in\reals^{V}$, and the return equals the reward $\gamma={u(\hat{\boldsymbol{v}}_{\mathrm{GCN}})}/{u(\hat{\boldsymbol{v}}_{\mathrm{Gr}})}$.
Our objective is to find the optimal set of parameters $\bbXi$ that maximizes the expected return for \emph{network state} $(\ccalG,\bbS, \bbu)$ drawn from a target distribution~$\Omega$
\vspace{-0.1in}
\begin{subequations}\label{E:GDPG}
\begin{align}
    \bbXi^{*} & = \argmax_{\bbXi \in \reals^{|\bbXi|}} J(\bbXi)\;, \\
    \text{s.t. }  J(\bbXi) & = \mathbb{E}_{(\ccalG,\bbS, \bbu)\sim\Omega}\left[ \gamma(\ccalG, \bbu, \bbz) \right]\;,\label{E:GDPG:obj}\\
    \gamma(\ccalG, \bbu, \bbz) & = u(\hat{\boldsymbol{v}}_{\mathrm{GCN}})/u(\hat{\boldsymbol{v}}_{\mathrm{Gr}})\;,\label{E:GDPG:reward}\\
    \bbz &=\Psi_{\ccalG}(\bbS;\bbXi)\;, \\
    \hat{\boldsymbol{v}}_{\mathrm{GCN}} &=g_d(\ccalG, \bbu\odot\bbz)\;,
    \hat{\boldsymbol{v}}_{\mathrm{Gr}} = g_c(\ccalG, \bbu)\;.\label{E:GDPG:sch}
\end{align}
\end{subequations}
Notice that functions $g_c(\cdot)$ and $g_d(\cdot)$ are respectively the efficient CGS and LGS as detailed in Section~\ref{sec:problem:greedy}, thus circumventing the need to exactly solve the MWIS problem.
According to the deterministic policy gradient theorem \cite[e.q. (9)]{silver2014deterministic}, the gradient of $J(\bbXi)$ can be found by 
\begin{equation}\label{E:dpg_grad}
\begin{split}
    \nabla J(\bbXi) &= \mathbb{E}_{S\sim\Omega_S}\left[\nabla_{\bbXi} \Psi_{\ccalG}(\bbS;\bbXi) \nabla_{\bbz} Q(S,\bbz)\right] \\
    &\approx \mathbb{E}_{S\sim\Omega_S}\left[ \nabla \Psi_{\ccalG}(\bbS;\bbXi)\bbQ(S,\bbz)\right]\;.
\end{split}
\end{equation}
In \eqref{E:dpg_grad}, vector $\bbQ(S,\bbz) \in\reals^{V}$ approximates the contribution of each dimension of action $\bbz$ to the Q-value  $Q(S,\bbz)=\mathbb{E}_{\bbu\sim\Omega_{u}^{S}}\left[\gamma(\ccalG, \bbu, \bbz)\right]$, $\Omega_S$ is the distribution of state $S=(\ccalG,\bbS)$, and $\Omega_{u}^{S}$ is the conditional distribution of $\bbu$ under state $S$.
Indeed, if the Q-value $Q(S,\bbz)$ were to be a differentiable function of the action $\bbz$, we would have $\bbQ(S,\bbz) = \nabla_{\bbz} Q(S,\bbz)$. 
However, since $Q(S,\bbz)$ is non-differentiable due to the non-differentiable function $g_d(\cdot)$ in \eqref{E:GDPG:sch}, we propose a proxy for the true credit assignment vector of action $\bbz$
\begin{equation}\label{E:q_vec}
    \bbQ(S,\bbz) \!=\! \mathbb{E}_{\bbu\sim\Omega_{u}^{S}}\left[ \gamma(\ccalG, \bbu, \bbz){\hat{\bbv}}  \right],\; 
    \hat{\bbv}\!=\! \left[\mathbbm{1}_{\hat{\boldsymbol{v}}_{\mathrm{GCN}}}(v) | v\in\ccalV\right],
\end{equation}
where $\hat{\bbv}\in\{0,1\}^{V}$ is the indicator vector of $\hat{\boldsymbol{v}}_{\mathrm{GCN}}$.
The intuition behind \eqref{E:q_vec} is that the more likely a link $v\in\ccalV$ is scheduled under state-action pair $(S,\bbz)$, the more it contributes to $Q(S,\bbz)$.
Based on \eqref{E:dpg_grad} and \eqref{E:q_vec} and a learning rate $\alpha\in(0,1)$, we can update the parameters $\bbXi$ through the following stochastic gradient ascent
\begin{equation}\label{E:stochastic_grad}
    \bbXi \leftarrow \bbXi+\alpha\widehat{\nabla J(\bbXi)},\;\widehat{\nabla J(\bbXi)} = \gamma(\ccalG, \bbu, \bbz)\nabla \Psi_{\ccalG}(\bbS;\bbXi) {\hat{\bbv}}\;.
\end{equation}

In training, we draw a batch of $\ccalG^{(i)}$ from the training dataset and $\bbu_v^{(i)}\in\mathbb{U}(0,1)$, collect the tuple $<\ccalG^{(i)},\bbS^{(i)},\bbu^{(i)}, \bbz^{(i)}, \hat{\boldsymbol{v}}_{\mathrm{GCN}}^{(i)}, \gamma^{(i)}>$ of each forward pass $i$, and then update the GCN by \eqref{E:stochastic_grad}.
{Intuitively, \eqref{E:stochastic_grad} encourages the GCN to generate the solution $\hat{\boldsymbol{v}}_{\mathrm{GCN}}^{(i)}$ for input $ (\ccalG^{(i)}, \bbu^{(i)}) $ by an amount proportional to its quality $\gamma^{(i)}$. 
Therefore, the aggregate effect of \eqref{E:stochastic_grad} on a batch of $ (\ccalG^{(i)}, \bbu^{(i)}) $ is moving $\bbXi$ towards generating solutions of larger $\gamma$.}
To prevent overfitting, after each update of $\bbXi$, the GCN-LGS is tested on a small independent validation dataset drawn from the target distribution~$\Omega$,
and  the $\bbXi$ that yields the best $\gamma$ on the validation dataset is kept as $\bbXi^{*} $.

\section{Numerical experiments}
\label{sec:results}

The performance of the GCN-based MWIS solvers is evaluated on synthetic random graphs and as schedulers in wireless networks. 
The comparative baselines are CGS, LGS~\cite{joo2012local}, and message passing (MP)~\cite{paschalidis2015message}.
To evaluate the contribution of the GCNs in our solvers, we also replace the GCNs in our solvers by random values \cite{bother2022s} ($\bbz_v\in\mathbb{N}(1,0.2)$, denoted as Random-LGS and Random-CRS) and a local 5-layer perceptron taking node degree as input and trained in the same way as GCN (denoted as MLP(5)-LGS and MLP(5)-CRS).
Threshold local greedy~\cite{joo2015distributed} and Ising \cite{Li18Ising} are not included since they perform worse than LGS/CGS.
The quality of an approximate solution $\hat{\boldsymbol{v}}$ is evaluated by its approximation ratio (AR)  $u(\hat{\boldsymbol{v}})/u(\boldsymbol{v}^{*})$, where the optimal solution $\boldsymbol{v}^*$ is obtained by solving the computationally expensive integer programming formulation of MWIS \cite{sanghavi2009message,paschalidis2015message} using the Gurobi solver~\cite{gurobi}.

The synthetic conflict graphs for training and testing are generated from the Erdős–Rényi (ER)~\cite{erdds1959random} and Barabási–Albert (BA)~\cite{Albert02} models. 
The ER model is completely determined by two parameters: the number of vertices $V$, and the probability of edge-appearance $p$.
The BA model is also determined by two parameters: the number of vertices $V$ and the number of edges, $m$, that each new vertex forms during the preferential attachment process.
In the experiments, we set $m=Vp=\bar{d}$ so that graphs from the ER and BA models have the same expected average degree.
By default, the vertex utilities are drawn following a uniform distribution $u(v)\sim\mathbb{U}(0,1)$.

\begin{table}[!t]
	\centering
	\caption{Mean ARs of solvers on synthetic graphs 
	}\vspace{-0.05in}
	\label{tab:result:random}
	\renewcommand{\arraystretch}{1.0}
	\footnotesize
	\begin{tabular}{l|c|c|c|c}
		Solver & Train & ER set & BA set & Thpt. \\ \hline
		CGS, LGS [Joo12] & - & $0.897$ & $0.858$ & $0.921$ \\
		MP [Paschalidis15] & - & $0.907$ & $0.892$ & - \\
		GCN(20)-CRTS & ER & $0.989$ & $0.993$ & - \\
		GCN(5)-CGS & ER & $0.976$ & $0.972$ & $0.960$ \\ \hline
		GCN(1)-CRS-v & ER & $0.978$ & $0.979$ & $0.995$ \\ 
		GCN(1)-CRS-e & ER & $0.985$ & $0.986$ & $0.996$ \\ 
		{Random-CRS} & - & $0.952$ & $0.943$ & - \\
		{MLP(5)-CRS} & ER & $0.970$ & $0.969$ & - \\ \hline
		GCN(1)-LGS & ER & $0.932$ & $0.937$ & $0.954$ \\
		GCN(1)-LGS-it & ER & $0.936$ & $0.942$ & $0.956$ \\
		{Random-LGS} & - & $0.873$ & $0.837$ & - \\
		{MLP(5)-LGS} & ER & $0.917$ & $0.902$ & - \\ \hline
	\end{tabular}
	\vspace{-0.2in}
\end{table}

\begin{figure*}[t]
	\centering
	\vspace{-0.2in}
	\subfloat[]{
		\includegraphics[width=0.48\linewidth]{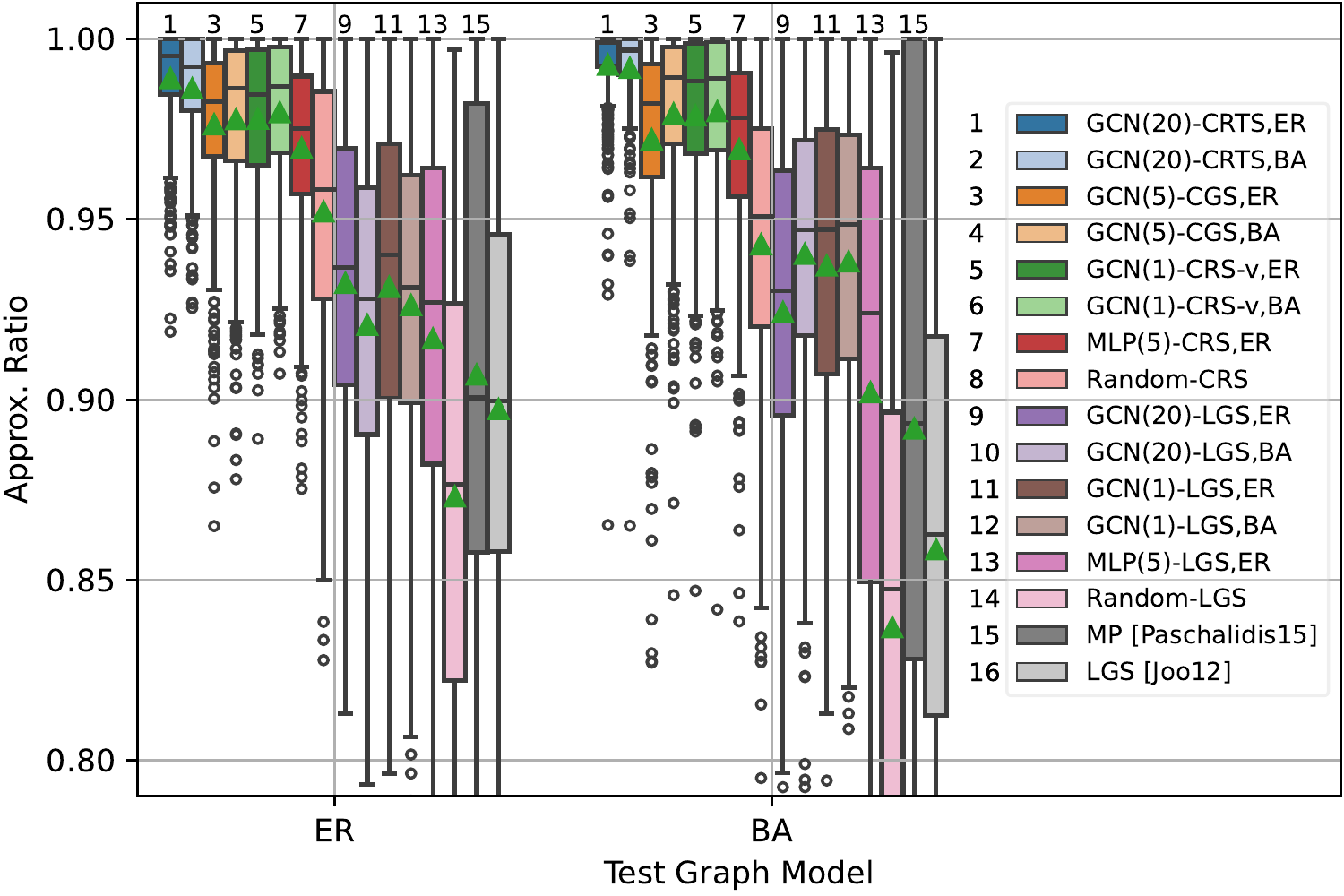}    
		\label{fig:generalization}
	}
	\subfloat[]{
		\includegraphics[width=0.48\linewidth]{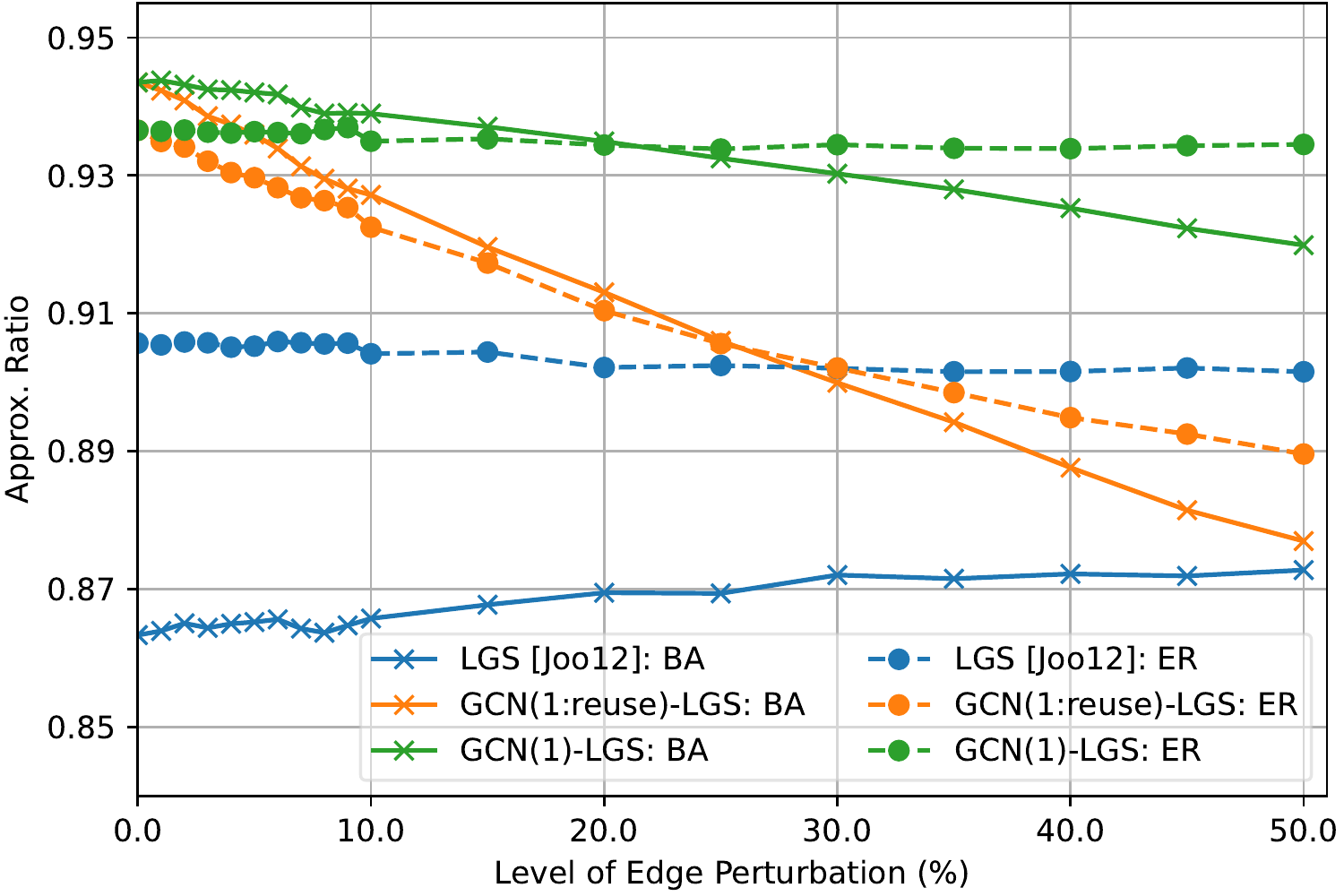}
		\label{fig:topology}   
	}
	\vspace{-0.05in}
	\caption{(a) Boxplot of approximation ratios of GCN-based MWIS solvers trained on ER and BA models and tested on both settings.
		(b) Approximation ratios of reusing topological embeddings under topology mismatches.}
	\vspace{-0.15in}
	
\end{figure*}

The hyperparameters of the evaluated GCNs are as follows: the numbers of layers $L\in\{1,5,20\}$, the size of every hidden layer is $g_l=32$ for $L>1$. 
The $L$-layered GCN is denoted as GCN($L$).
{To balance exploration and computational efficiency, the branching factor is selected as $B=32$, the same as in  \cite{li2018combinatorial}.}
Each GCN is trained on a set of 5900 random graphs drawn from an ER model unless otherwise specified. 
The training set comprises 5000 graphs of size $V\in \left\{100, 150, 200, 250, 300\right\}$ and expected average degree $\bar{d}\in \left\{2, 5, 7.5, 10, 12.5\right\}$ ($200$ graphs per $(V, \bar{d})$), and 900 graphs of size $V\in\left\{30,100\right\}$ and edge probability $p\in\left\{0.1,0.2,\dots,0.9\right\}$ ($50$ graphs per $(V, p)$).
The timeout of GCN-CRTS is set to 5 minutes.
The GCN in GCN-CRTS is trained by supervised learning with a maximum of 200 epochs as explained in Section~II-D of~\cite{supplement}. 
The GCNs in GCN-CRS and GCN-LGS are trained as described in Section~\ref{sec:train}, for which the settings include a batch size of 200 for experience replay, 25 epochs, and periodic gradient reset.\footnote{Training typically takes 30 minutes on a workstation with a specification of 16GB memory, 8 cores and Geforce GTX 1070 GPU. The source code is published at \url{https://github.com/zhongyuanzhao/distgcn}}
This configuration is used for the training of GCN-CGS as described in Section~III of~\cite{supplement}.

\subsection{Performance on Synthetic Random Graphs}\label{sec:results:random}

\subsubsection{Results on ER graphs}
First, the GCN-based MWIS solvers, baseline heuristics, and optimal solver are tested on a set of 500 ER graphs of size $V\in \left\{100, 150, 200, 250, 300\right\}$ and average degree $\bar{d}\in \left\{2, 5, 10, 15, 20\right\}$, with $20$ instances for each pair of $(V, \bar{d})$. 
The average ARs of the tested solvers are listed in Table~\ref{tab:result:random} (under the `ER set' column) for the entire test set, and illustrated in Figs.~\ref{fig:cmpx:degree} and~\ref{fig:cmpx:size} as a function of the average degree and size of the tested graphs, respectively.
{Note that the same trained GCN model of GCN(1) is used by solvers of GCN(1)-CRS-v, GCN(1)-CRS-e, GCN(1)-LGS, GCN(1)-LGS-it, and GCN(1)-LGS-$N$ for $N=3$ and $N=4$.}
The relative performance of all the tested solvers decreases on larger and denser graphs.
Intuitively, the size of the MWIS decreases as the graph becomes denser (the average degree increases), hence a wrong selection of a vertex incurs a higher cost in the AR.
Also, the MWIS problem is harder on larger graphs due to the larger search space.

Among the centralized solvers, the GCN-based centralized solvers outperform the vanilla CGS on average AR by a gap of $7.7\%$ to $9.2\%$. 
The GCN(20)-CRTS achieves the top average AR of $0.989$, but is also most sensitive to the graph size (see Fig.~\ref{fig:cmpx:size}), since the larger search space lowers its chance of finding a good solution before timeout.
Compared to GCN(20)-CRTS, our proposed GCN(1)-CRS-v reduces its runtime by at least two orders of magnitude (i.e., from 5 minutes to up to a few seconds), at the cost of $1.2\%$ average AR, and is less sensitive to graph size.
It demonstrates that rollout search can reach a good solution quickly with the same trained model.
Note that the performance of rollout search is largely determined by its guiding heuristic.
With the guiding heuristic of enhanced CGS described in Section~\ref{sec:search:rollout}, the AR of GCN(1)-CRS-e can be boosted by an average of $0.8\%$ from that of GCN(1)-CRS-v.
{On the ER test set, the GCN(5)-CGS performance is similar to that of the GCN(1)-CRS-v with slight advantage on denser and larger graphs.
It shows that GCN can match the vanilla CGS in estimating the Q-values when the test set matches the training set.}
Moreover, GCN(5)-CGS further reduces the runtime by an order of magnitude on those worst-case instances (i.e., from seconds to sub-seconds)\footnote{For deployment, these presented runtimes can be further optimized.} and has the lowest sensitivity to graph size. 
{In comparison, the runtime of the highly optimized exact Gurobi solver on the worst-case instances is 4 hours.}

Our proposed GCN-guided distributed solvers, GCN(1)-LGS and GCN(1)-LGS-it, also outperform the baselines of LGS and MP. 
Prepending a 1-layer GCN improves LGS by $3.5\%$ on average AR, and this difference is more conspicuous (close to $5\%$) in the more challenging case of denser graphs.
Although MP can find the optimal solution on some sparsely connected graphs (i.e., small average degree), it becomes virtually the same as LGS when the average degree increases to the range of typical conflict graphs in practice (i.e., average degree of 10 to 20).
The GCN(1)-LGS-it outperforms GCN(1)-LGS by $0.4\%$ on average, and its advantage is mostly on larger and denser graphs, as feeding the residual graph to GCN(1) improves the consistency. 
On the other hand, the LGS-3, LGS-4, GCN(1)-LGS-3, and GCN(1)-LGS-4 achieve average ARs of $0.883$, $0.896$, $0.923$, and $0.931$, respectively. 
From the curves of these truncated solvers in Fig.~\ref{fig:cmpx}, we can find that improperly truncating the LGS can severely degrade the performance on larger and denser graphs, on which LGS requires more iterations to complete. 
As shown in \cite{zhao2021icassp}, increasing the number of layers $L$ from 1 to 3 does not yield better performance.
{Our GCN-based solvers performs notably better than their counterparts based on MLP(5) and random values, where MLP(5)-LGS outperforms vanilla LGS while Random-LGS does not. 
It shows that our training method allows both GCN and MLP to learn meaningful representations, and GCN is more representative than MLP.}

\begin{figure*}[t]
	\centering
	\vspace{-0.25in}
	\subfloat[]{
		\includegraphics[width=0.48\linewidth]{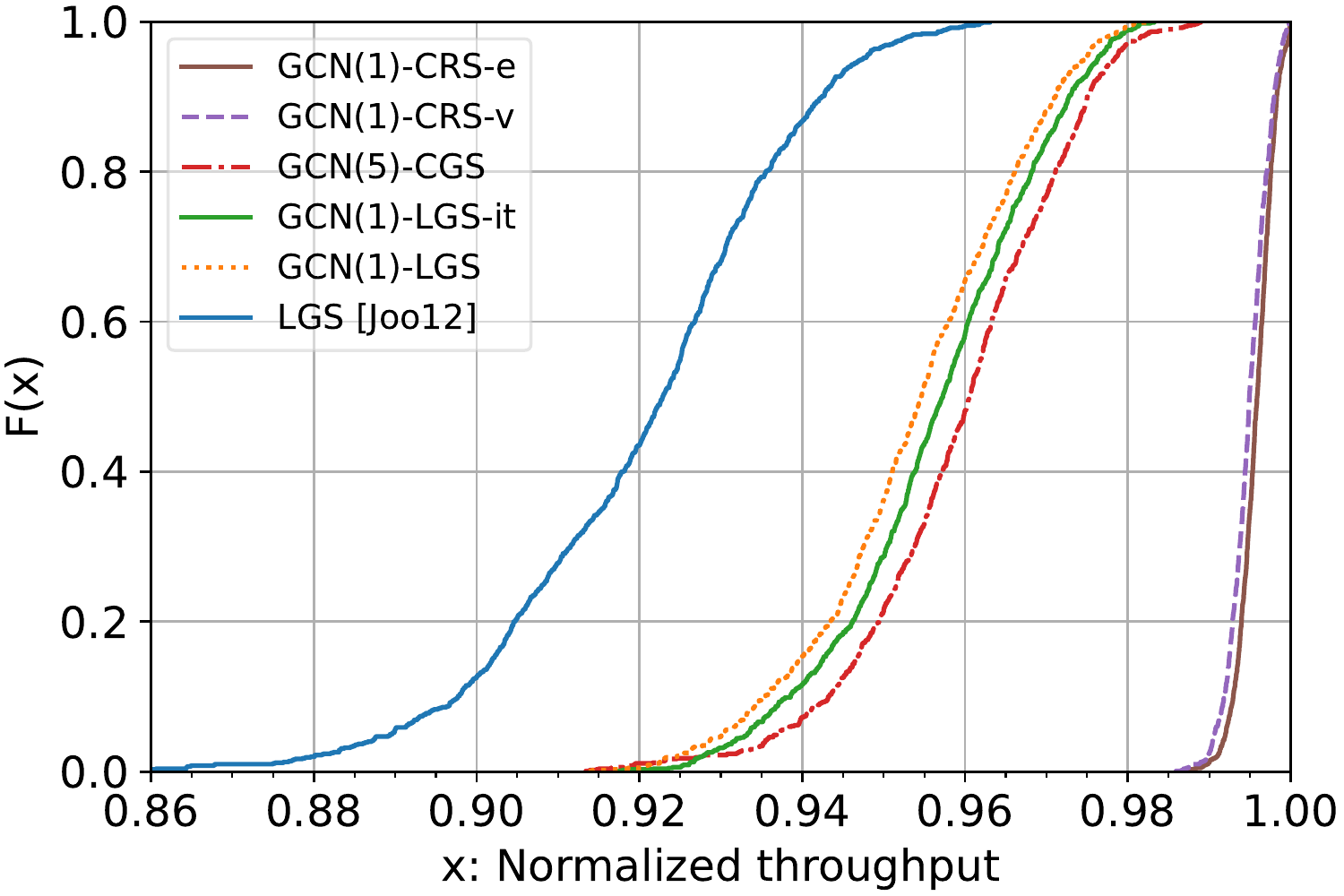}
		\label{fig:traffic}  
	}
	\subfloat[]{
		\includegraphics[width=0.48\linewidth]{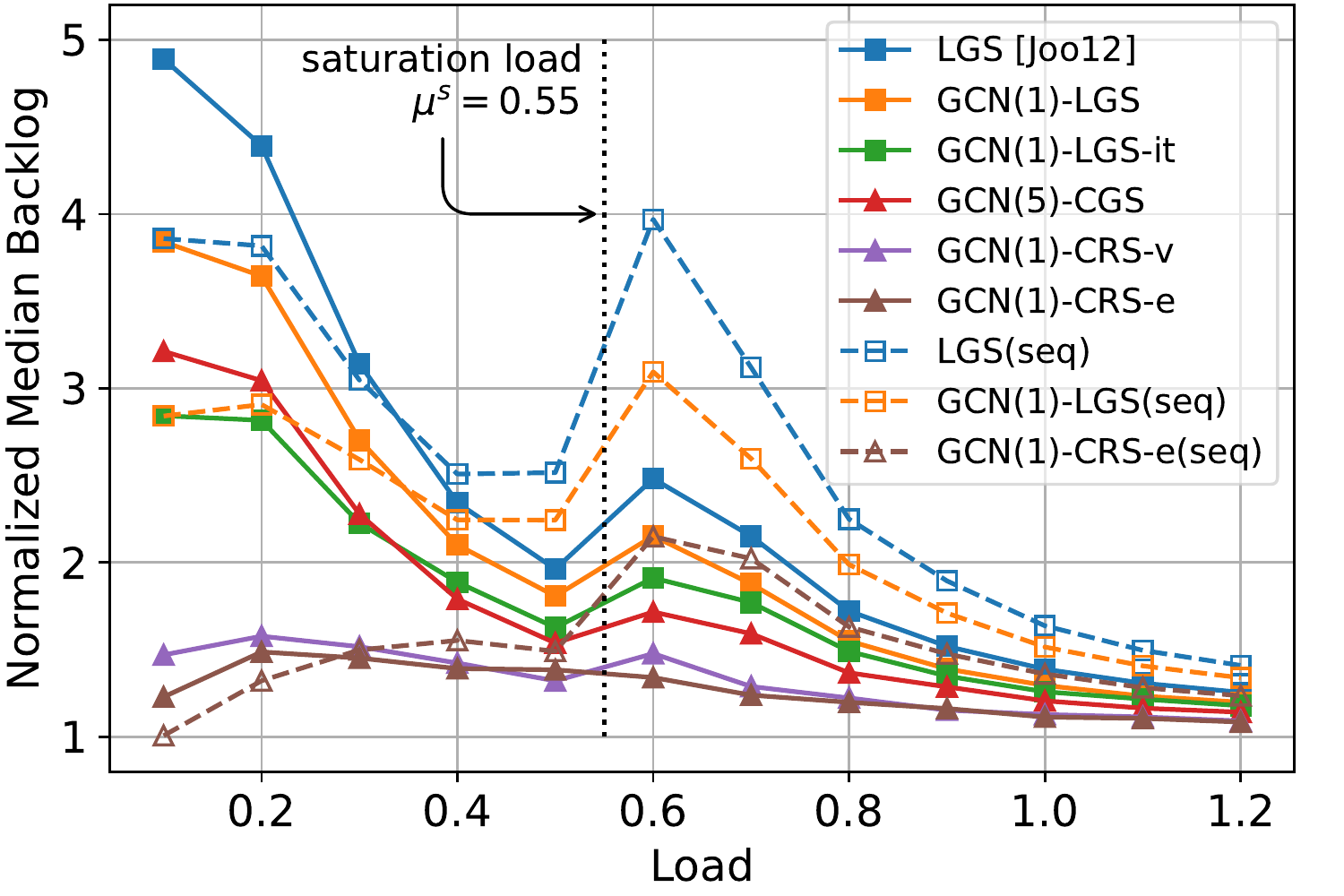}
		\label{fig:backlog:3ch} \label{fig:backlog}
	}   
	\vspace{-0.05in}
	\caption{Scheduling in single radio networks, 
		(a) eCDFs of the normalized throughput of the distributed and centralized schedulers w.r.t. the optimal scheduler, in single-channel scheduling under oversaturated traffic. Per-link utility $u(v)=\min(q(v),r(v))$. The average ARs of throughput of these schedulers are listed in Table~\ref{tab:result:random}. 
		(b) Normalized median backlog (smaller is better) of different schedulers w.r.t. the optimal scheduler in multichannel scheduling with 3 sub-channels under various network traffic loads. 
		Per-link utility $u(v)=q(v)r(v)$. 
	}   
	\vspace{-0.2in}
\end{figure*}

\subsubsection{Generalizability}
Next, we examine how well the GCN-based MWIS solvers generalize across graph models.
For a given set of hyperparameters, two versions of GCN-based solvers are trained on two different training sets generated from ER and BA models, respectively, with other configurations identical to the default setting. 
Each version of this solver is then tested on the ER test set of the previous experiment and a BA test set with identical configuration ($m=Vp=\bar{d}$).
The ARs of the proposed and baseline solvers on these two test sets are illustrated as mean values in Table~\ref{tab:result:random}, and as box plots in Fig.~\ref{fig:generalization}, where means are marked by green triangles.

Given the heavy-tailed degree distribution of BA graphs~\cite{Albert02}, the topology-agnostic solvers of LGS and MP experience drops of $3.9\%$ and $1.5\%$  in AR, respectively.
In contrast, GCN-based MWIS solvers, regardless of training set, achieve relatively consistent performance on both ER and BA test sets, e.g. an increase of $0.1\%$ to $0.7\%$ in AR on BA graphs, and outperform LGS and MP by a greater margin on BA graphs, as shown in Table~\ref{tab:result:random} and Fig.~\ref{fig:generalization}, which underscores the value of taking topology into account.
In general, GCN-based MWIS solvers tend to perform slighty better if they are trained on the same graph model of the test set, showing good transferability across graph models. 
To understand the impact of $L$, GCNs of 1 layer and 20 layers are evaluated for GCN-LGS.
Deeper GCNs do not significantly improve the mean ARs, but tend to present smaller variance compared to shallower GCNs. 
Moreover, deep GCNs are more tuned to the training graph model. 
For example, GCN($20$)-LGS attains the best performance when trained and tested in BA, but underperforms compared to GCN(1)-LGS when trained on one graph model and tested on the other. 
A similar pattern can be found in GCN-based centralized solvers, the solver trained and tested on the same graph model has a performance $0.5\%$ greater than when trained and tested on different graph models.
Results in Section~\ref{sec:results:throughput} further shows good transferability across per-link utility distributions.

\subsubsection{Topology mismatch} 
Lastly, we measure the robustness of the reusable topological node embedding $\bbz(\ccalG)$ (introduced in Section~\ref{sec:solution:reuse}) against topology changes in wireless networks, which can be attributed to mobility and shadowing. 
The reusing strategy, denoted as GCN(1:reuse)-LGS, applies the topological embedding $\bbz^{\ccalG^{(0)}}$ generated from a baseline graph $\ccalG^{(0)}$ to 100 instances of similar graphs $\ccalG^{(i)}, i\in\{1,\dots,100\}$ generated by replacing each edge in $\ccalG^{(0)}$ with a random new edge with a given probability denoted as the level of edge perturbation (e.g., the x-axis of Fig.~\ref{fig:topology}),
and each $\ccalG^{(i)}$ is associated with a new realization of random utility $\bbu^{(i)}\sim\mathbb{U}(\mathbf{0},\mathbf{1})$.
The tested baseline graphs are 500 random graphs generated by ER or BA models with $V\in\{100,150\}$ and $m=\bar{d}\in\{2,5,10,15,20\}$. 
The ARs of GCN(1:reuse)-LGS versus the normalized edit distance are illustrated in Fig.~\ref{fig:topology}, along with the LGS and the GCN(1)-LGS using $\bbz^{\ccalG^{(i)}}$ for $\ccalG^{(i)}$, as the control group. 
On average, the reusable topological embedding $\bbz^{\ccalG^{(0)}}$ can stand up to $30\%$ of edge perturbation on ER graphs and $50\%$ on BA graphs before its gain over LGS diminishes. 
Note that the control group is also slightly influenced by the topological change, since the underlying degree distribution of the similar graphs generated by our method will shift towards the ER model.
Since the topology change of a wireless network is usually several orders of magnitude slower than the channel fading that defines the coherence time slot, this result shows that the reusing strategy detailed in Section~\ref{sec:solution:reuse} allows GCN-LGS to be implemented at the same local complexity of vanilla LGS.

\subsection{GCN-based Throughput-Optimal Scheduling}
\label{sec:results:throughput}

To understand how the performance of the proposed solvers on synthetic random graphs could be transferred to scheduling, throughput-optimal scheduling in wireless ad-hoc networks is simulated. 
These networks consist of $100$ users randomly located in a square of area $250$. 
A link is established if the distance between two users is smaller than $1$, and two links interfere with each other if they have incident users within distance of $4$. 
A $1$-hop flow with random direction is created on each link. 
The exogenous arriving packets at each source node follow a Poisson arrival with a prescribed arrival rate $\lambda$.
The link rate $r(v)$, defined as the number of packets that can be transmitted through link $v$ in a time slot, is drawn independently from a normal distribution $\mathbb{N}(50, 25)$  clipped to $\left[0,100\right]$,  
to capture a clipped rectified linear function of signal-to-noise ratio (SNR)  caused by the effects of fading, constant transmit power, and lognormal shadowing. 
The network traffic load is defined as $\mu=\mathbb{E}(r)/\lambda$, and the {saturation load}, $\mu^{s}$, is defined as the load that makes the average queue length under optimal scheduling to equal the average link rate, $\mathbb{E}(q|{\mu^{s}})=\mathbb{E}(r)$. 
A total of $100$ realizations of random wireless networks are generated, typically with $40$-$60$ links, and the average degrees of their conflict graphs range from $7.7$ to $26.9$ with a mean of $13.2$.
For each conflict graph, $10$ scheduling instances of $200$ time slots are executed by each tested scheduler.
In each scheduling instance, the same realization of arrivals and link rates is used for each scheduler.

\subsubsection{Oversaturated traffic}
We first consider a single-radio single-channel scenario, where the exogenous packets arrive at an oversaturated rate, and the per-link utility is the number of packets the link can deliver, i.e., $u(v)=\min(q(v),r(v))$, where $q(v)$ is the backlog of link $v$. 
In this case, the average throughout {is identical} to the average utility a scheduler achieves per time slot.
{The average normalized throughputs achieved by the tested schedulers w.r.t. the optimal scheduler are listed in Table~\ref{tab:result:random} (Thpt.), and
their corresponding empirical CDFs are illustrated in Fig.~\ref{fig:traffic}, showing consistent improvements over LGS by GCN-based schedulers.} 
In oversaturated traffic, GCN(1)-LGS closes the suboptimality gap of LGS by $42\%$ on average, while the centralized schedulers GCN(1)-CRS-e and GCN(1)-CRS-v are near optimal. 

\subsubsection{Varying traffic}
Next, we consider the scenarios of a single-radio wireless network with three sub-channels, in which the traffic load varies from $0.1$ to $1.2$ with the saturation load $\mu^{(s)}=0.55$, and the per link utility function is $u(v)=q(v)r(v)$ \cite{tassiulas1992,Lin06,lin2007distributed,bhandari2010scheduling}. 
{By default, multi-channel scheduling is based on a single multi-channel conflict graph.}
To simulate frequency diversity, the multi-channel conflict graph is generated by adding an edge ($v_1^k$, $v_2^k$) for $k\in{1,2,3}$ at a probability of {$80\%$} if edge ($v_1$, $v_2$) appears in the single-channel conflict graph of the first scenario.
Compared to single channel scheduling, the size and density of the 3-channel conflict graph are both tripled, which would substantially increase the suboptimality gap of the heuristics, as shown in Fig.~\ref{fig:cmpx}.
Note that higher traffic load reduces the coefficient of variation (relative standard deviation) of per-link utility across the network, further enlarging the suboptimality gap.
{The normalized latencies of the tested heuristic schedulers w.r.t. the optimal scheduler, measured by normalized median backlog (smaller is better) by traffic load,} are presented in Fig.~\ref{fig:backlog}.
The overall ranking of the 6 tested heuristics on median backlog in 3-channel scheduling is the same as that on throughput in single channel scheduling.
In the stable region, the median backlog of LGS decreases from $4.9$ times of the optimal scheduler under lightweight traffic, to $1.95$ times under near-saturation traffic. 
Our distributed schedulers, GCN(1)-LGS and GCN(1)-LGS-it, can respectively close the suboptimality gap of LGS by $22\%$ to $10\%$ and $54\%$ to $32\%$ between lightweight and near-saturation traffics.
Our centralized schedulers, GCN(1)-CRS-v and GCN(1)-CRS-e, only increase the median backlog of the optimal scheduler by a maximum of $50\%$ and $60\%$, respectively.
{In addition, multi-channel schedulers based on solving a sequence of MWIS problems on sub-channels, denoted by dashed curves and postfix '(seq)' in the legends in Fig.~\ref{fig:backlog}, outperform their counterparts on normalized latency only under light traffic loads, e.g., $\mu=0.1,0.2$.}

In both single-channel and 3-channel scheduling, centralized scheduler GCN(5)-CGS performs only slightly better than GCN(1)-LGS-it, in contrast with its good performance on synthetic random graphs.
This result shows that our GCN-based rollout search can generalize well to unseen graph size, graph density, and weight distribution, which is crucial for scalable scheduling in wireless multihop networks, whereas a naive integration of GCN and deep Q learning could not.

\vspace{-0.1in}
\subsection{Concluding Remarks on Numerical Experiments}

The numerical results show that a minimal GCN (GCN(1)) with only 2 trainable parameters [e.g., $\bbTheta_0^1,\bbTheta_1^1\in\reals$ in~\eqref{eq:1layer}], when integrated into various algorithmic frameworks, can substantially close the suboptimality gap of the baseline greedy heuristics for the MWIS problem. 
The enhancement in an efficient heuristic can be transferred to improved efficiency or performance of more sophisticated exact and approximate solvers, which usually employ efficient heuristics as intermediate steps or to obtain warm-start solutions.

In practice, our centralized solvers require efficient implementations that leverage the sparsity of the graph to cope with the computational complexity listed in Table~\ref{tab:complexity:central}.
The distributed solver GCN(1)-LGS can be implemented by increasing the size of control message without additional local exchanges.
Lastly, the advantage of GCN-based heuristics over the greedy baseline is more conspicuous for wireless networks that are denser, larger, and with more sub-channels.

\section{Conclusions and Future Work}
\label{sec:conclusions}

We proposed several graph-aware efficient MWIS solvers for link scheduling in wireless networks. 
The centralized solver uses a lightweight GCN to guide centralized rollout tree search, and can achieve near optimal performance on small- to middle-sized networks with a complexity of $\ccalO(V^2)$. 
By leveraging the topology-awareness and distributed nature of GCNs, the distributed solvers enhance the distributed greedy heuristic and retain its efficiency, and can achieve superior performance in larger and denser networks with a local communication complexity of $\ccalO(\log V)$. 
The GCN can be trained on simulated networks without exactly solving the NP-hard MWIS problem, and generalizes well across different types of graphs. 
Moreover, our approach is agnostic to the specific per-link utility, thus, it can be used in conjunction with many existing distributed scheduling protocols. 
{In practice, our approach could perform poorly due to corrupted parameters of the GCN, or large estimation errors in per-link utility, e.g., link rate.}
Future research efforts include: 
1)~Incorporating state-awareness into the per-link utility by taking into account the causal relationship between network state transition and scheduling decision, 
2)~Considering the scheduling problem for wireless networks operating on non-orthogonal channels, and
3)~Developing distributed online learning schemes to enable training in real world scenarios.


\bibliographystyle{IEEEtran}
\bibliography{strings,refs}

\begin{IEEEbiography}
	[{\includegraphics[width=1in,height=1.25in,clip,keepaspectratio]{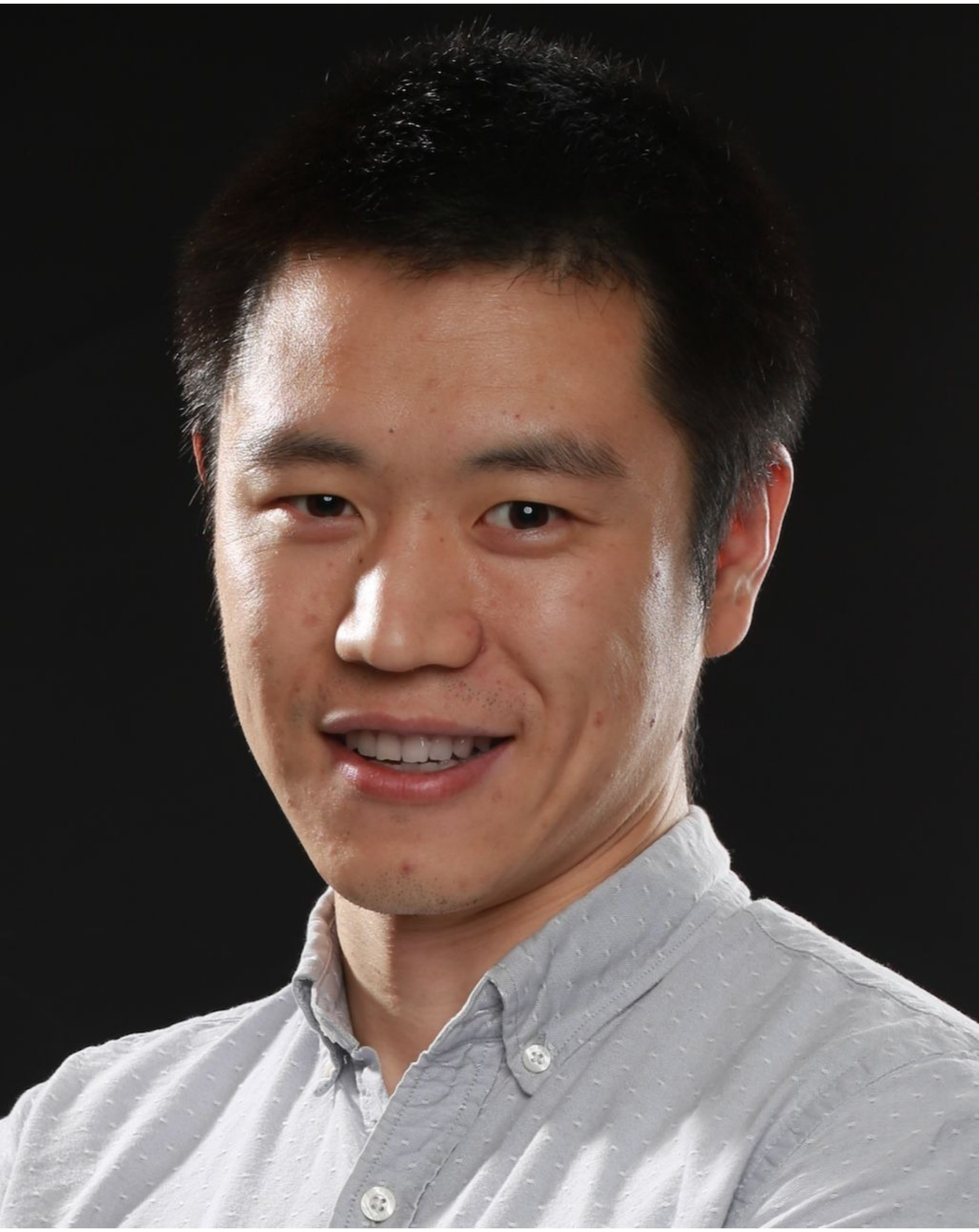}}]{Zhongyuan Zhao} (S'13--M'18) received his B.Sc. and M.S. degrees in Electronic Engineering from the University of Electronic Science and Technology of China, Chengdu, China, in 2006 and 2009, respectively. He received his Ph.D. degree in Computer Engineering from the University of Nebraska-Lincoln, Lincoln, NE, in 2019, under the guidance of Prof. Mehmet C. Vuran. From 2009 to 2013, he worked for ArrayComm and Ericsson, respectively, as an engineer developing 4G base-station. Currently, he is a postdoctoral research associate at the Department of Electrical and Computer Engineering of Rice University, advised by Prof. Santiago Segarra. Dr. Zhao’s current research interests focus on machine learning and signal processing for wireless communications and networking.
\end{IEEEbiography}
\vspace{-0.5in}

\begin{IEEEbiography}
	[{\includegraphics[width=1in,height=1.25in,clip,keepaspectratio]{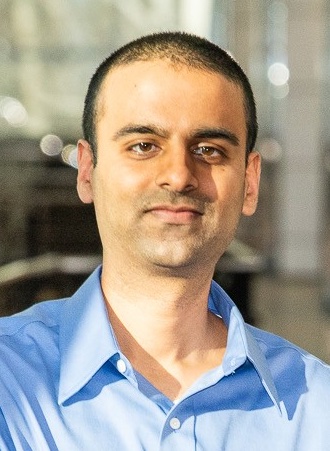}}]{Gunjan Verma}
	received the B.S. degree in mathematics, computer science, and economics from Rutgers University, the M.A. degree in computational biology from Duke University, and the M.S. degree in mathematics from Johns Hopkins University. He is currently a Computer Scientist with the U.S. Army Research Laboratory (ARL), Adelphi, MD, USA. His research interests include Bayesian statistics and machine learning and their application to problems in networking, communications, and robotics.
\end{IEEEbiography}
\vspace{-0.5in}


\begin{IEEEbiography}
	[{\includegraphics[width=1in,height=1.25in,clip,keepaspectratio]{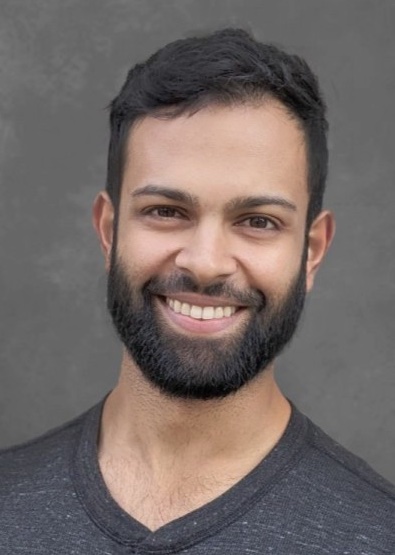}}]{Chirag Rao}
	received the B.S. degree in electrical and computer engineering (ECE) from Cornell University in 2013, and the M.S. degree in computer science from Johns Hopkins University in 2018.
	He is currently pursuing the Ph.D. degree with the Laboratory for Information and Decision Systems (LIDS), Massachusetts Institute of Technology. Since 2013, he has been a Researcher with the U.S. Army Research Laboratory. His research interests include wireless networking and multi-agent systems, leveraging applied probability, optimization, and machine learning to analyze and develop algorithms and protocols for wireless networks.
\end{IEEEbiography}
\vspace{-0.5in}

\begin{IEEEbiography}
	[{\includegraphics[width=1in,height=1.25in,clip,keepaspectratio]{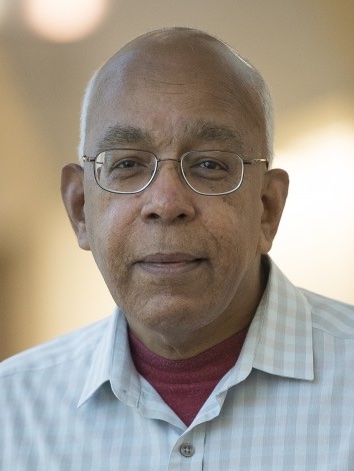}}]{Ananthram Swami}
	(Life Fellow, IEEE) received the B.Tech. degree from IIT-Bombay, the M.S. degree from Rice University, and the Ph.D. degree from the University of Southern California (USC), all in electrical engineering. He is currently with the U.S. Army's DEVCOM Army Research Laboratory (ARL) as the Army’s Senior Research Scientist (ST) for Network Science, and is an ARL fellow. Prior to joining ARL, he held positions with Unocal Corporation, USC, and CS-3. He has held visiting faculty positions at INP, Toulouse, and Imperial College, London.  His recent awards include the 2018 IEEE ComSoc MILCOM Technical Achievement Award and the 2017 Presidential Rank Award (Meritorious).
\end{IEEEbiography}
\vspace{-0.5in}

\begin{IEEEbiography}
	[{\includegraphics[width=1in,height=1.25in,clip,keepaspectratio]{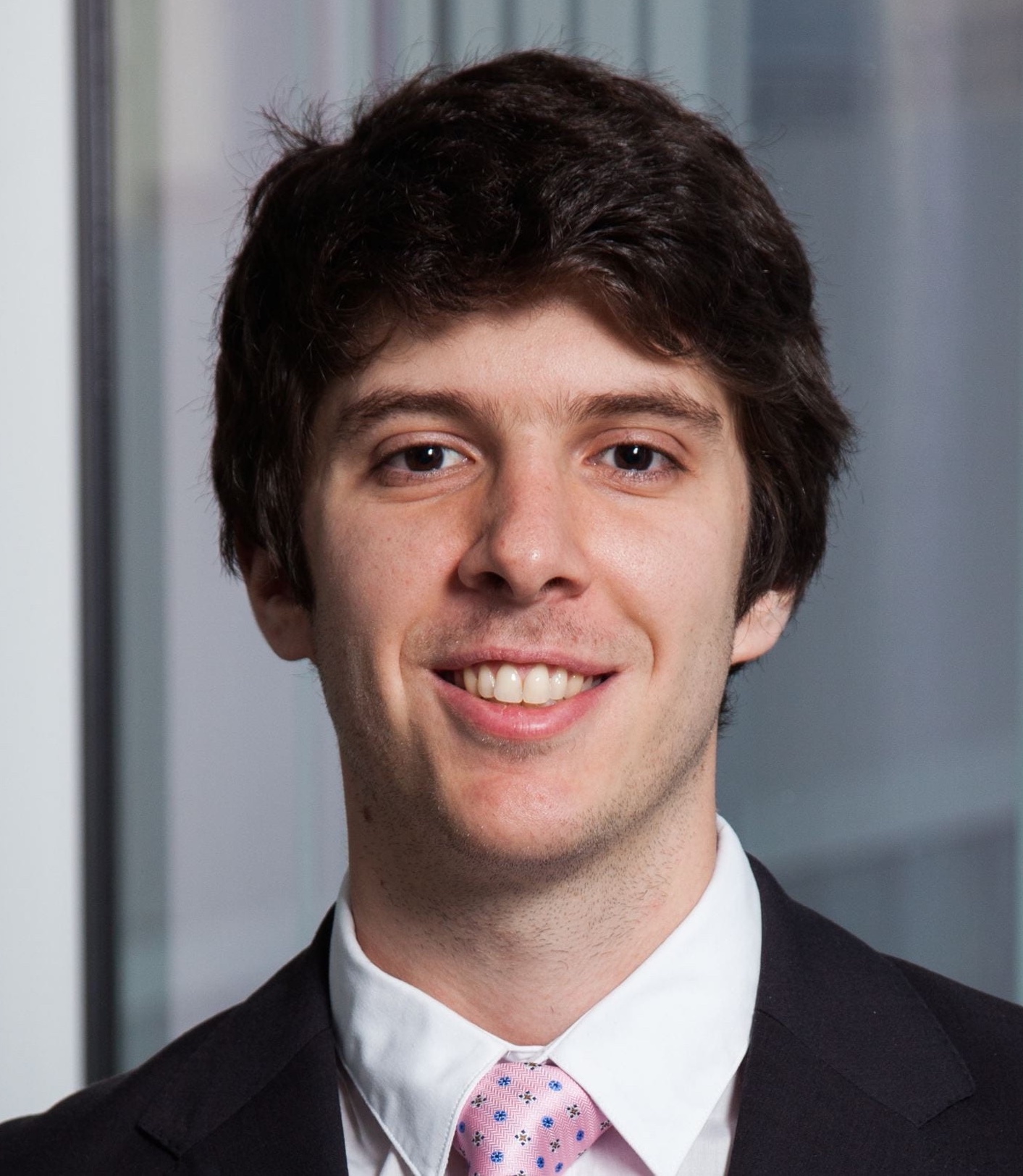}}]{Santiago Segarra} (Senior Member, IEEE) received the B.Sc. degree (Hons.) (Valedictorian) in industrial engineering from the Instituto Tecnológico de Buenos Aires (ITBA), Argentina, in 2011, the M.Sc. in electrical engineering from the University of Pennsylvania (Penn), Philadelphia, in 2014 and the Ph.D. degree in electrical and systems engineering from Penn in 2016. From September 2016 to June 2018 he was a postdoctoral research associate with the Institute for Data, Systems, and Society at the Massachusetts Institute of Technology. Since July 2018, Dr. Segarra has been an Assistant Professor in the Department of Electrical and Computer Engineering at Rice University. His research interests include network theory, data analysis, machine learning, and graph signal processing. He received the ITBA’s 2011 Best Undergraduate Thesis Award in Industrial Engineering, the 2011 Outstanding Graduate Award granted by the National Academy of Engineering of Argentina, the 2017 Penn’s Joseph and Rosaline Wolf Award for Best Doctoral Dissertation in Electrical and Systems Engineering, the 2020 IEEE Signal Processing Society Young Author Best Paper Award, the 2021 Rice’s School of Engineering Research + Teaching Excellence Award, and five best conference paper awards.
\end{IEEEbiography}

\end{document}


\markboth{Supplemental Materials for Zhao \MakeLowercase{\textit{et al.}}: Link Scheduling using Graph Neural Networks}{Supplemental Materials for Zhao \MakeLowercase{\textit{et al.}}: Link Scheduling using Graph Neural Networks}

\maketitle

\section{Centralized greedy solver}
A centralized greedy solver (\emph{CGS}), denoted by function $\hat{\boldsymbol{v}}_{\mathrm{Gr}} = g_c(\ccalG,\bbu)$, builds the approximate solution $\hat{\boldsymbol{v}}_{\mathrm{Gr}}$ to the MWIS problem in an iterative fashion by first adding to $\hat{\boldsymbol{v}}_{\mathrm{Gr}}$ the vertex with the largest utility, deleting its neighbors as potential candidates, and repeating this procedure until all vertices are either added to $\hat{\boldsymbol{v}}_{\mathrm{Gr}}$ or deleted, as detailed in Algorithm~\ref{algo:cgs} \cite{dimakis2006sufficient}. 

\begin{algorithm}
\caption{{Centralized greedy solver $\hat{\boldsymbol{v}}_{\mathrm{Gr}} = g_c(\ccalG,\bbu)$}}
\label{algo:cgs}
\hspace*{\algorithmicindent} \textbf{Input:} $\ccalG, \bbu$ \\
\hspace*{\algorithmicindent} \textbf{Output:} $\hat{\boldsymbol{v}}_{\mathrm{Gr}}$ 
\begin{algorithmic}[1] 
\STATE $\hat{\boldsymbol{v}}_{\mathrm{Gr}}\gets\varnothing$; $\ccalG'(\ccalV',\ccalE')\gets\ccalG(\ccalV, \ccalE)$
\WHILE{$\ccalG' \neq \phi$}
\STATE $
    v = \underset{v_i \in \ccalV'}{\argmax }\, u(v_i)
    $
\STATE $\hat{\boldsymbol{v}}_{\mathrm{Gr}} \gets \hat{\boldsymbol{v}}_{\mathrm{Gr}}\cup \{v\}$
\STATE $ \ccalG'\gets\ccalG' \setminus (\hat{\boldsymbol{v}}_{\mathrm{Gr}}\cup\mathcal{N}(\hat{\boldsymbol{v}}_{\mathrm{Gr}})) $
\ENDWHILE
\end{algorithmic}
\end{algorithm}\vspace{-0.1in}

\section{GCN-guided centralized random tree search} \label{sec:search}
The GCN-guided centralized random tree search (\emph{GCN-CRTS}) is modified from a similar solver of the unweighted MIS problem \cite{li2018combinatorial}.
The basic idea of GCN-CRTS is to reach as many random terminal nodes as possible in a given time interval to increase its chance of finding a good solution from them.

\subsection{Iterative Framework}\label{sec:search:framework}
 
\begin{figure}[t]
	\centering
	\ifpaper
    \subfloat[]{
            \includegraphics[height=2.1in]{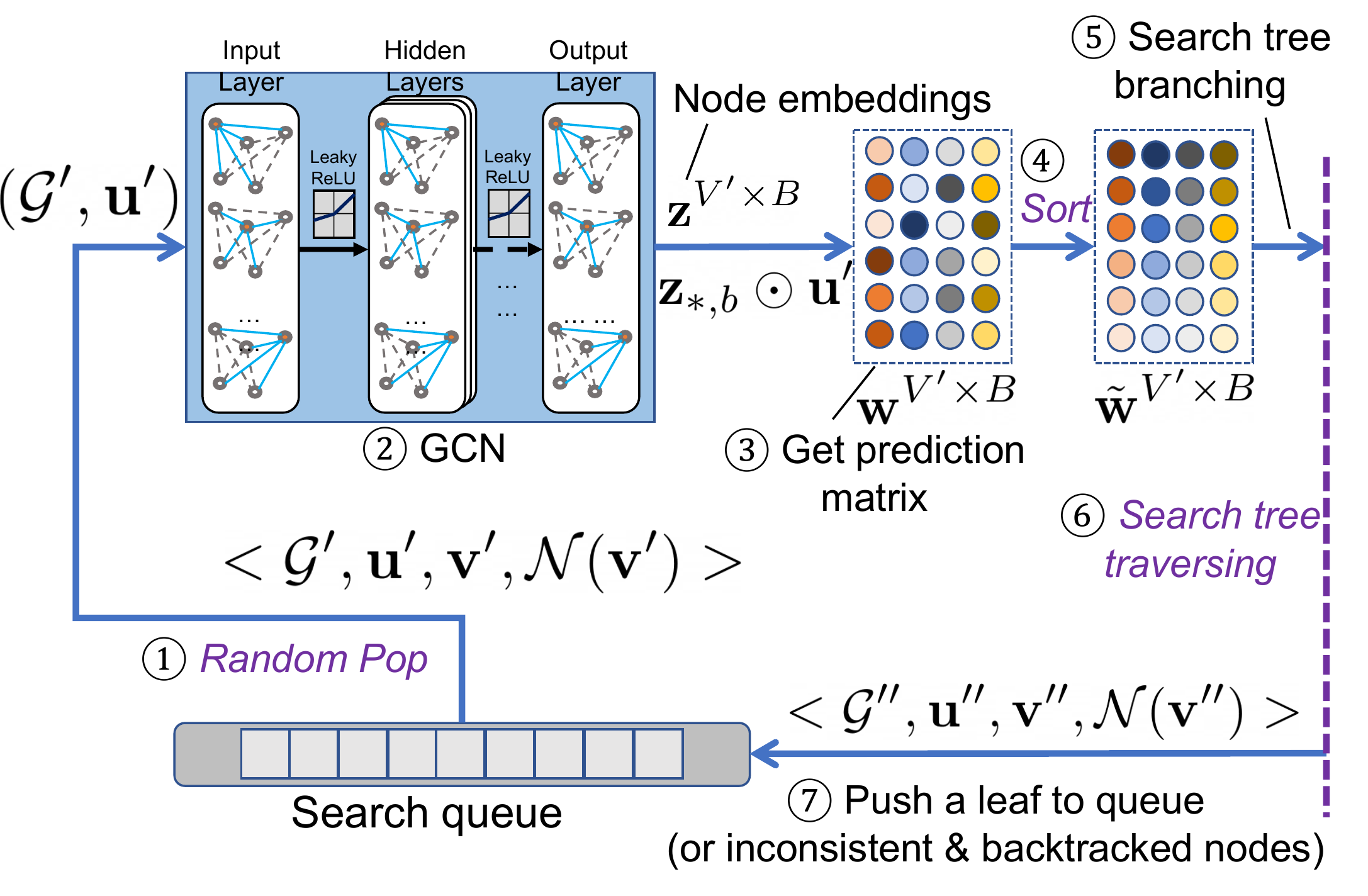}
            \label{fig:gcncentral}
    }  \hfil
   \subfloat[]{
            \includegraphics[height=2.1in]{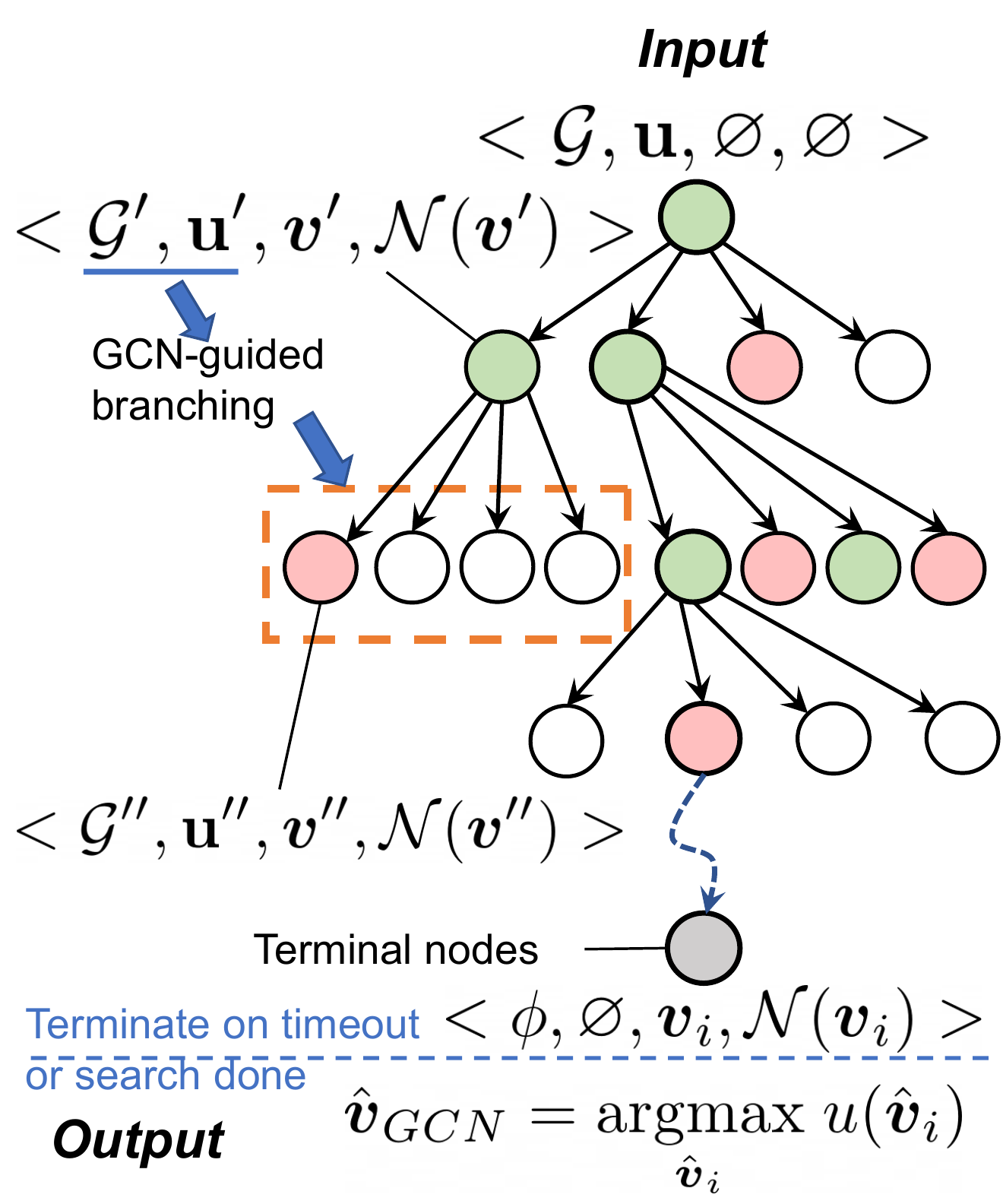}
            \label{fig:treesearch}
    }\hfil
    \fi
	\caption{Architecture of GCN-based centralized MWIS solvers. 
	(a)~Iterative framework of GCN-guided tree search. 
	(b)~Random tree search with timeout, where the consistent and inconsistent nodes are marked by green and red, respectively.}
	\label{fig:dqnsearch}
\end{figure}

Compared to GCN-CRS in the main manuscript \cite{zhao2022twc}, the GCN-CRTS has a similar iterative algorithmic framework, as illustrated in Fig.~\ref{fig:dqnsearch}, and the same definition of search tree.
The possible transitions of the state of each iteration form a search tree, as illustrated in Fig~\ref{fig:treesearch}. 
The solver finds an approximate solution by traversing from the root of the search tree to a terminal state.

{On initialization, the root node is pushed into the search queue.}
In each iteration, the solver expands a non-terminal node and proceeds to a child node in 7 steps, as illustrated in Fig.~\ref{fig:gcncentral}. 
In step 1, a non-terminal node $<\ccalG',\bbu',\boldsymbol{v}',\ccalN(\boldsymbol{v}')>$ is randomly popped from the search queue, of which the enclosed residual graph $\ccalG'$ and residual utility vector $\bbu'$ are fed into the GCN.
In step 2, based on the inputs of $\ccalG'$ and $\bbu'$, the GCN generates the node embedding matrix $\bbZ' \in \left[0,1\right]^{V'\times B}$, which collects the topology-aware scaling factor $\bbZ'_{v,b}=z'_{b}(v)$ for all $v \in \ccalV'$ and $b\in\{1,\dots,B\}$, as $\bbZ' = \Psi_{\ccalG'}(\bbu'; \bbXi)$, where $B$ is the branching factor of the search tree configured to create a large search tree with a large amount of distinct terminal nodes, $\Psi_{\ccalG'}$ is an $L$-layered GCN defined on the graph $\ccalG'$, $\bbu'$ is the residual utility vector, and $\bbXi$ represents the collection of trainable parameters of the GCN.
In step 3, a prediction matrix $\bbW'\in\reals^{V' \times B}$ is created from $\bbZ'$ and $\bbu'$, of which an element $\bbW'_{v,b}=w'_{b}(v)=z'_{b}(v)\cdot u'(v)$ represents the topology-aware utility of vertex $v$. 
In step 4, each column of the prediction matrix is sorted in descending order.
More precisely, denoting by $\bbX_{*, i}$ the $i$th column of the generic matrix $\bbX$, we obtain the sorted prediction matrix $\tilde{\bbW}'_{*,b}=\mathrm{sort}(\bbW'_{*,b})$ and the sorted vertex matrix $\tilde{\bbV}'_{*,b}=\mathrm{argsort}(\bbW'_{*,b})$, as illustrated in Fig.~\ref{fig:gcncentral} where elements in $\bbW'$ and $\tilde{\bbW}'$ with larger values are represented by darker colors.
Next, in step 5, up to $B$ prediction vertices are extracted from the sorted vertex matrix $\tilde{\bbV}'$ to create the $B$ branches (next states) of the current node (current state) in the search tree, respectively. 
The child node $<\ccalG'',\bbu'',\boldsymbol{v}'',\ccalN(\boldsymbol{v}'')>$ on a branch is generated based on one of the $B$ prediction vertices, denoted as $v$, as follows: 
\begin{equation}\label{eq:child}
\ccalG''=\ccalG'\backslash (v\cup\ccalN(v))\;,
\boldsymbol{v}''=\boldsymbol{v}'\cup\{v\}\;.
\end{equation}
In step 6, the solver selects one of the $B$ children nodes to proceed according to the traversing strategy detailed in Sections~\ref{sec:search:random}.
Lastly, in step 7, if the selected child node (next state) is a non-terminal node, it is pushed into the search queue as a new leaf in the search tree.

\subsection{Graph Convolutional Network Design}\label{sec:search:gcn} 

The GCN in GCN-CRTS has the same input and hidden layers as the GCN in the main manuscript \cite{zhao2022twc}, and a slightly different output structure. 
The output layer of GCN has a dimension of $g_{L}=2B$, with sigmoid/softmax activation applied to $B$ pairs of logits. 
In each pair of logits for vertex $v$, the first element is the probability that $v\in\boldsymbol{v}^{*}$ and the second element is probability that $v\notin\boldsymbol{v}^{*}$, where $\boldsymbol{v}^{*}$ is an MWIS for input $(\ccalG',\bbu')$.
The node embedding matrix $\bbZ' \in \left[0,1\right]^{V'\times B}$ is obtained by extracting the first elements of the $B$ pairs of logits for all $v\in\ccalV'$.
The GCN is trained by supervised learning, as detailed in Section~\ref{sec:search:random}. 

\subsection{Random Tree Search} \label{sec:search:random}

At step 6 of each algorithmic iteration described in Section~\ref{sec:search:framework}, the GCN-CRTS traverses the search tree as follows:
First, a column $b\in\{1,\dots,B\}$ of the sorted vertex matrix $\tilde{\bbV}'$ is selected uniformly at random.
Then, a depth first search is performed by testing the vertices in $\tilde{\bbV}'_{*,b}$ one by one according to \eqref{eq:child}, until reaching a terminal node (gray node in Fig.~\ref{fig:treesearch}) or an inconsistent node (pink nodes in Fig.~\ref{fig:treesearch}).
If the tested vertex $v\in\ccalN\left(\boldsymbol{v}''\right)$, the node $<\ccalG'',\bbu'',\boldsymbol{v}'',\ccalN(\boldsymbol{v}'')>$ is marked as inconsistent and pushed to the search queue in Fig.~\ref{fig:dqnsearch}.
To ensure that all terminal nodes could be reached, a backtracking mechanism is employed so that each consistent node can also be pushed to the search queue with some prescribed probability. 
The search is terminated on timeout or the condition of all terminal nodes being reached, and the output is the best solution observed throughout the tree search
\begin{equation}
    \hat{\boldsymbol{v}}_{GCN} = \argmax_{\hat{\boldsymbol{v}}_i}u(\hat{\boldsymbol{v}}_i)\;,
\end{equation}
where $\hat{\boldsymbol{v}}_i$ is the solution from the $i$th terminal node being reached.
The GCN-CRTS is detailed in Algorithm~\ref{algo:gcn-crts}.
To improve the quality of the solution, the search tree can be traversed by multiple parallel threads (i.e., parallel for loop start from line 6 in Algorithm~\ref{algo:gcn-crts}) that share the same search queue to reach more terminal nodes before timeout.

\begin{algorithm}[t!]
\caption{{GCN-guided Centralized Random Tree Search}}
\label{algo:gcn-crts}
\hspace*{\algorithmicindent} \textbf{Input}: $\ccalG, \bbu$ \\
\hspace*{\algorithmicindent} \textbf{Output}: $ \hat{\boldsymbol{v}}_{\mathrm{GCN}} $
\begin{algorithmic}[1] 
\REQUIRE $ B, p_{b}\in\left[0,1\right] $, MaxRunTime
\STATE Initialize search queue $\ccalQ \gets \{<\ccalG,\bbu,\varnothing,\varnothing>\}$
\STATE Initialize set of candidate solutions $\ccalC \gets \varnothing$
\WHILE{ $\ccalQ = \varnothing $ or Timeout }
\STATE $<\ccalG',\bbu',\boldsymbol{v}',\ccalN(\boldsymbol{v}')> = \mathrm{RandomPop}(\ccalQ)$ 
\STATE $\bbZ' = \Psi_{\ccalG'}(\bbu'; \bbXi)$ 
\FORALL{$ b \in \{1,\dots, B\} $ }
\STATE $\bbW'_{*,b}=\bbZ'_{*,b}\odot\bbu'$ 
\STATE $\tilde{\bbV}'_{*,b} = \mathrm{ArgSortDescending}(\bbW'_{*,b})$ 
\STATE $<\ccalG'',\bbu'',\boldsymbol{v}'',\ccalN(\boldsymbol{v}'')>\gets <\ccalG',\bbu',\boldsymbol{v}',\ccalN(\boldsymbol{v}')>$
\FOR{$ j = 1,\dots, |\ccalV'| $ }
\STATE $v = \tilde{\bbV}'_{j, b}$ 
\IF{ $ v \notin \ccalN(\boldsymbol{v}'') $ }
\STATE $\ccalG''\gets \ccalG'\backslash (\{v\}\cup\ccalN(v))\;,\boldsymbol{v}''\gets\boldsymbol{v}'\cup\{v\}\;$ 
\STATE Draw $p \in \mathbb{U}(0,1)\;$, stop $\gets$ False 
\ELSE
\STATE $\ccalG''\gets\ccalG'\backslash \{v\}\;,\boldsymbol{v}''\gets\boldsymbol{v}'\;$ 
\STATE $p \gets 1 $, stop $\gets$ True
\ENDIF
\IF{ $ \ccalG''\neq \phi $ and $p<p_b$ }
\STATE $\ccalQ\gets\ccalQ\cup\{<\ccalG'',\bbu'',\boldsymbol{v}'',\ccalN(\boldsymbol{v}'')>\}$ 
\ELSE 
\STATE $\ccalC\gets \ccalC\cup\{\boldsymbol{v}''\}$
\ENDIF
\IF{stop}
\STATE break;
\ENDIF
\ENDFOR 
\ENDFOR 
\ENDWHILE
\STATE $\hat{\boldsymbol{v}}_{GCN} = \underset{\hat{\boldsymbol{v}}\in\ccalC}{\argmax}\,u(\hat{\boldsymbol{v}})\;$
\end{algorithmic}
\end{algorithm}

\subsection{Training}
The GCN in GCN-CRTS is trained by supervised learning with a loss function \cite{li2018combinatorial}
\begin{equation}
	\ell(\bbXi; \ccalG, \bbu, \bby) = \min_{b=1}^{B} \frac{1}{V}\sum_{v=1}^{V}\ccalC(\bby_v,\bbZ_{v,b})\;,
\end{equation}
where $\bby\in\{0,1\}^{V}$ is the label vector ($\bby_v=1$ if $v\in \boldsymbol{v}^{*}$ and $0$ otherwise) and $\ccalC(y,z)=-\left[y\log(z)+(1-y)\log(1-z)\right]$ is the cross entropy function. 
Synthetic random graphs are used as the training data. 
Since obtaining the optimal solution is NP-hard, the label vector $\bby$ is generated by heuristics for computational efficiency. 
Specifically, the labels are generated by selecting the best solution from two guiding heuristics: linear programming \cite{KaKo2009} and centralized greedy solver.
Thanks to the iterative framework and nature of GCN, the GCN-CRTS can generalize and outperform the training graphs and labels in terms of size \cite{li2018combinatorial} and quality.
However, the GCN-CRTS will take a relatively long time to find a good solution (e.g., several minutes), therefore is not suitable for link scheduling. 

\section{GCN-guided Centralized Greedy Search}  \label{sec:search:greedy}

Compared to GCN-CRTS, GCN-guided centralized greedy search (\emph{GCN-CGS}) can reach a good terminal node quickly, thus suitable for link scheduling.
The GCN-CGS is modified from the deep Q network in \cite{khalil2017learning} by replacing the Node2Vec module with a GCN, with implementation detailed in Algorithm~\ref{algo:gcn-cgs} and explained as follows.

{The iterative framework of GCN-CGS is almost identical to that of GCN-CRS with only two exceptions:
1) The branching factor $B=|\ccalV'|$ is no longer a hyperparameter.
2) Instead of $\bbw'$, the node embedding $\bbz'\in\reals^{V'}$ is directly used as the vector of Q values of the action space.
} 
For a non-terminal node $<\ccalG',\bbu',\boldsymbol{v}',\ccalN(\boldsymbol{v}')>$ in the search tree, the observation of the agent is $x'=(\ccalG',\bbu')$, where $\ccalG'$ is the residual graph, $\bbu'$ is the corresponding utility vector, and the action space $\ccalA'$ of the agent is the vertex set $\ccalV'$ of the graph $\ccalG'$. 
With the input $(\ccalG',\bbu')$, the GCN generates the node embedding $\bbz'$ as the Q values of the action space, based on which the action is selected by an $\epsilon$-greedy policy. 
The next state $x''=(\ccalG'',\bbu'')$ is generated according to \eqref{eq:child}.
The search terminates upon reaching a terminal node.

The GCN is trained by finite-horizon reinforcement learning with a discount rate of $1.0$. 
The reward is defined as $\gamma=0$ for a non-terminal state, and $\gamma={u(\hat{\boldsymbol{v}}_{\mathrm{GCN}})}/{u(\hat{\boldsymbol{v}}_{\mathrm{Gr}})}$ for a terminal state, where $\hat{\boldsymbol{v}}_{\mathrm{Gr}}$ is obtained by the vanilla centralized greedy solver. 
In each iteration, the tuple of $<x',\ccalA',\gamma,x'',done>$ is saved to the replay buffer of the agent, where $done$ is a boolean variable indicating if $x'$ is on a terminal node. 
The GCN is trained by standard experience replay as used in deep Q learning.
The exploration rate $\epsilon$ is initialized to $1$ and decays exponentially during training, and set to $0$ for testing and deployment.
The GCN-CGS is enabled by the fact that a GCN can have state and action space of any dimensions.

\begin{algorithm}[t!]
\caption{{GCN-guided Centralized Greedy Search}}
\label{algo:gcn-cgs}
\hspace*{\algorithmicindent} \textbf{Input}: $\ccalG, \bbu$ \\
\hspace*{\algorithmicindent} \textbf{Output}: $\hat{\boldsymbol{v}}_{\mathrm{GCN}}$ 
\begin{algorithmic}[1] 
\STATE Initialize search queue $\ccalQ \gets \{<\ccalG,\bbu,\varnothing,\varnothing>\}$
\WHILE{ $\ccalQ = \varnothing $ }
\STATE $<\ccalG',\bbu',\boldsymbol{v}',\ccalN(\boldsymbol{v}')> = \mathrm{RandomPop}(\ccalQ)$ 
\STATE $\bbz' = \Psi_{\ccalG'}(\bbu'; \bbXi)$ 
\STATE $v = \underset{v_i\in\ccalV'}{\argmax}\,z'(v_i)\;$
\STATE $\ccalG''=\ccalG'\backslash (v\cup\ccalN(v))\;,\boldsymbol{v}''=\boldsymbol{v}'\cup\{v\}\;$ 
\IF{ $ \ccalG''\neq \phi $ }
\STATE $\ccalQ\gets\ccalQ\cup\{<\ccalG'',\bbu'',\boldsymbol{v}'',\ccalN(\boldsymbol{v}'')>\}$ 
\ELSE 
\STATE $\hat{\boldsymbol{v}}_{\mathrm{GCN}}\gets \boldsymbol{v}''$
\ENDIF
\ENDWHILE
\end{algorithmic}
\end{algorithm}

\section{Distributed Scheduler: GCN-LGS-it}
The \emph{GCN-LGS-it} solver, as detailed in Algorithm~\ref{algo:lgs-it}, can further improve the performance over the baseline GCN-LGS at the cost of higher local communication complexity.
In GCN-LGS-it, the GCN is placed before each inner iteration of LGS (lines 3-4 in Algorithm~\ref{algo:lgs-it}), so that the input of the LGS iteration $\bbw'=\bbz'\odot\bbu'$, where $\bbz'=\Psi_{\ccalG'}(\bbu';\bbXi)$, is based on the residual graph $(\ccalG',\bbS')$ from the previous iteration, rather than the input graph $(\ccalG,\bbS)$ as in the baseline GCN-LGS.

\begin{algorithm}[t!]
\caption{{GCN-LGS-it}}
\label{algo:lgs-it}
\hspace*{\algorithmicindent} \textbf{Input:} $\ccalG, \bbS, \bbu$ \\
\hspace*{\algorithmicindent} \textbf{Output:} $\hat{\boldsymbol{v}}_{\mathrm{GCN}}$ 
\begin{algorithmic}[1] 
\STATE $\hat{\boldsymbol{v}}_{\mathrm{GCN}}\gets\varnothing$; $\ccalG'(\ccalV',\ccalE')\gets\ccalG(\ccalV, \ccalE)$; $\bbm=\boldsymbol{0}$
\WHILE{$\ccalG' \neq \phi$}
\STATE $\bbz' = \Psi_{\ccalG'}(\bbS'; \bbXi)$
\STATE $\bbw' = \bbu'\odot\bbz'$
\FORALL{ $v\in \ccalV'$ }
\STATE $v$ exchanges $w'(v)$ with its neighbors $\forall v_i \in \ccalN_{\ccalG'}(v)$
\IF{ $ w'(v) > \underset{v_i \in \mathcal{N}_{\ccalG'}(v)}{\max}\, w'(v_i) $ } 
\STATE $m(v) \gets +1$, $v$ broadcasts a control message
\STATE $m({v_i})\gets -1, \forall v_i \in \mathcal{N}_{\ccalG'}(v)$
\ENDIF
\ENDFOR
\STATE $\hat{\boldsymbol{v}}_{\mathrm{Gr}} \gets \{v | v\in\ccalV', m(v)=1\}$
\STATE $\ccalV'\gets \{ v | v\in \ccalV', m(v)=0\}$, update $\ccalG'$ with new $\ccalV'$
\ENDWHILE
\end{algorithmic}
\end{algorithm}

\section{Construction of conflict graph}

The construction of the conflict graph (i.e., $\ccalG = (\ccalV,\ccalE)$) is a process separated from scheduling.
The scheduler needs to find an independent set per time slot, whereas the frequency of reconstructing or updating the conflict graph depends on the mobility, e.g., a conflict graph can be reused for many time slots in the scenarios of static or low-mobility.

A conflict graph can be constructed by a parallel process with a time complexity of $\ccalO(\Delta)$, where $\Delta$ is the maximum degree of the conflict graph $\ccalG$.
Moreover, the construction of conflict graph can be further blended into other processes, such as initial access, handover, and regular data communications. 
Essentially, a conflict graph can be constructed by each link learning who are its immediate interfering neighbors through channel monitoring.
In centralized scheduling, each link also needs to report its list of neighbors to the centralized scheduler (e.g., a base-station).
In practice, a conflict graph is often constructed and maintained in an incremental manner, for example, when a new link is formed (e.g., by a new or roaming device), it first learns its immediate neighbors, and then broadcasts its own ID to its neighbors during initial access or handover. 
Similarly, existing links maintain their list of neighbors by removing inactive links and adding new links in the background.
In such incremental mode, the conflict graph is maintained in parallel in the background.
Even if a conflict graph needs to be built from scratch, it only takes one round of information exchange for each link to learn its immediate neighbors, which has a time complexity of $\ccalO(\Delta)$.

In distributed scheduling, the whole conflict graph is not required, since each vertex (link) only needs the knowledge of its own neighborhood, e.g., in the distributed implementations of both GCN and LGS. 
In the case of centralized scheduling, where the whole conflict graph is required by the centralized scheduler running on a base-station, all links report their lists of neighbors to the base-station through control channel(s), which could also be implemented in an incremental manner by only reporting the changes.
At the base-station, the computational complexity of constructing a conflict graph from scratch based on the reported neighboring relationships is $\ccalO(|\ccalE|)$.
When the conflict graph is updated and maintained incrementally, the computational complexity is linear to the number of changes, $\ccalO(|\ccalE^{(t+1)} - \ccalE^{(t)}|)$. 
Indeed, the computational complexity of constructing a conflict graph is much lower than solving a MWIS instance with GCN-CRS.

In addition, the message exchanges between vertices and their neighbors are based on the assignment of control channels, which is part of the initial access and out of the scope of the scheduling problem.
The control channel assignment can be formulated as distributed graph coloring \cite{chung2017distributed}.
When there are more control channels than the maximum degree of the conflict graph $\Delta$, distributed graph coloring from scratch can be done in logarithmic rounds $\ccalO(\log^* V)$, and incremental (defective) coloring can be done in sublogarithmic rounds \cite{chung2017distributed}.
Such a low complexity ensures the scalability of distributed resource allocation.

Since constructing the conflict graph has a computational (local) complexity lower than the proposed centralized (distributed) MWIS solvers, and can be blended into other networking processes, we exclude the complexity of constructing a conflict graph from the complexity of schedulers.

\begin{figure}[t]
    \centering
    \vspace{-0.1in}
    \includegraphics[width=0.9\linewidth]{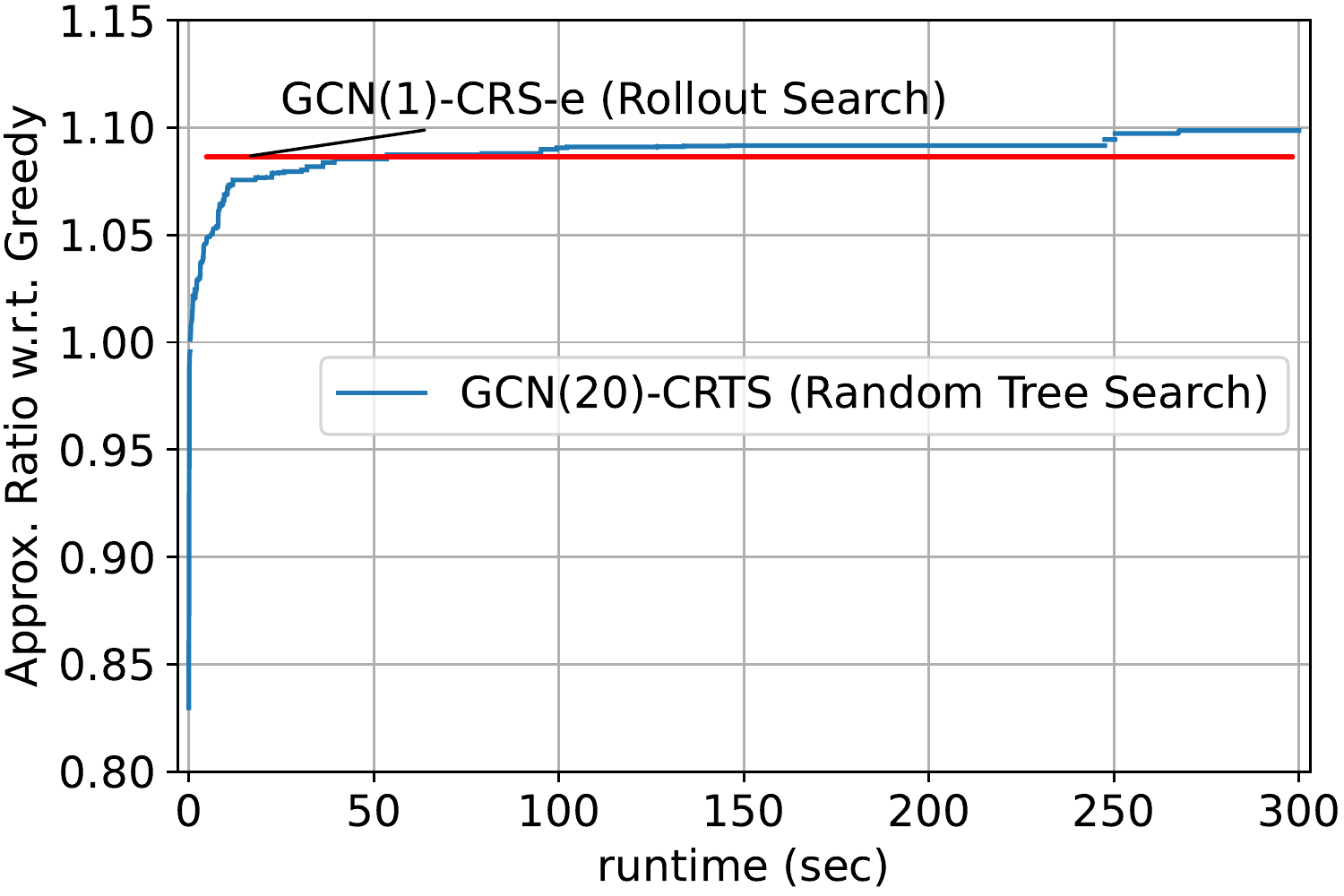}
    \caption{The evolution of average approximation ratio of GCN(20)-CRTS, the weighted modification of \cite{li2018combinatorial}, w.r.t. greedy solver, over time v.s. GCN-CRS (denoted as Rollout), on 20 instances of ER graphs with $|\ccalV|=300$ and an average degree of 20. The starting point of the curve, e.g., $x=0$, is about $0.87$.}
    \label{fig:arbytime}
    \vspace{-0.2in}
\end{figure}

\section{Additional Numerical Results}

On top of the experiments in \cite{zhao2022twc}, we present additional results to further demonstrate: 1) the difference between our approach and closely related work \cite{li2018combinatorial}, and 2) the generalizability of our proposed GCN-based MWIS solvers on graphs that are larger and/or denser than the graphs the GCNs are trained on.

\subsection{Comparison between GCN-CRS and GCN-CRTS}

The GCN-guided tree search in \cite{li2018combinatorial} randomly traverses mthe search tree predicted by the GCN, for any times, in order to find a good solution from the hundreds of thousands of candidate solutions it reached.
In Fig.~\ref{fig:arbytime}, we present the evolution of the approximation ratio of the best solution found by GCN(20)-CRTS, w.r.t. the greedy solution, over its total runtime of 300 seconds, where the blue curve is the average value of 20 instances of ER graphs with size $|\ccalV|=300$ and average degree of $Np=20$.
Although the initial candidate solution of GCN(20)-CRTS is only $0.83$ of the greedy solution, it quickly outperforms the greedy solution. 
However, it takes GCN(20)-CRTS over 50 seconds to find a solution better than the first solution found by GCN(1)-CRS, which takes less than 3 seconds.
In general, the runtime of GCN(20)-CRTS needs to be an order of magnitude higher than the runtime of GCN-CRS, so that the former could outperform the latter.
As a result, GCN-CRTS is not suitable for link scheduling, in which an MWIS instance needs to be solved in sub-seconds.

\begin{figure}[t]
    \vspace{-0.1in}
    \centering
    \subfloat[]{
    \includegraphics[width=0.9\linewidth]{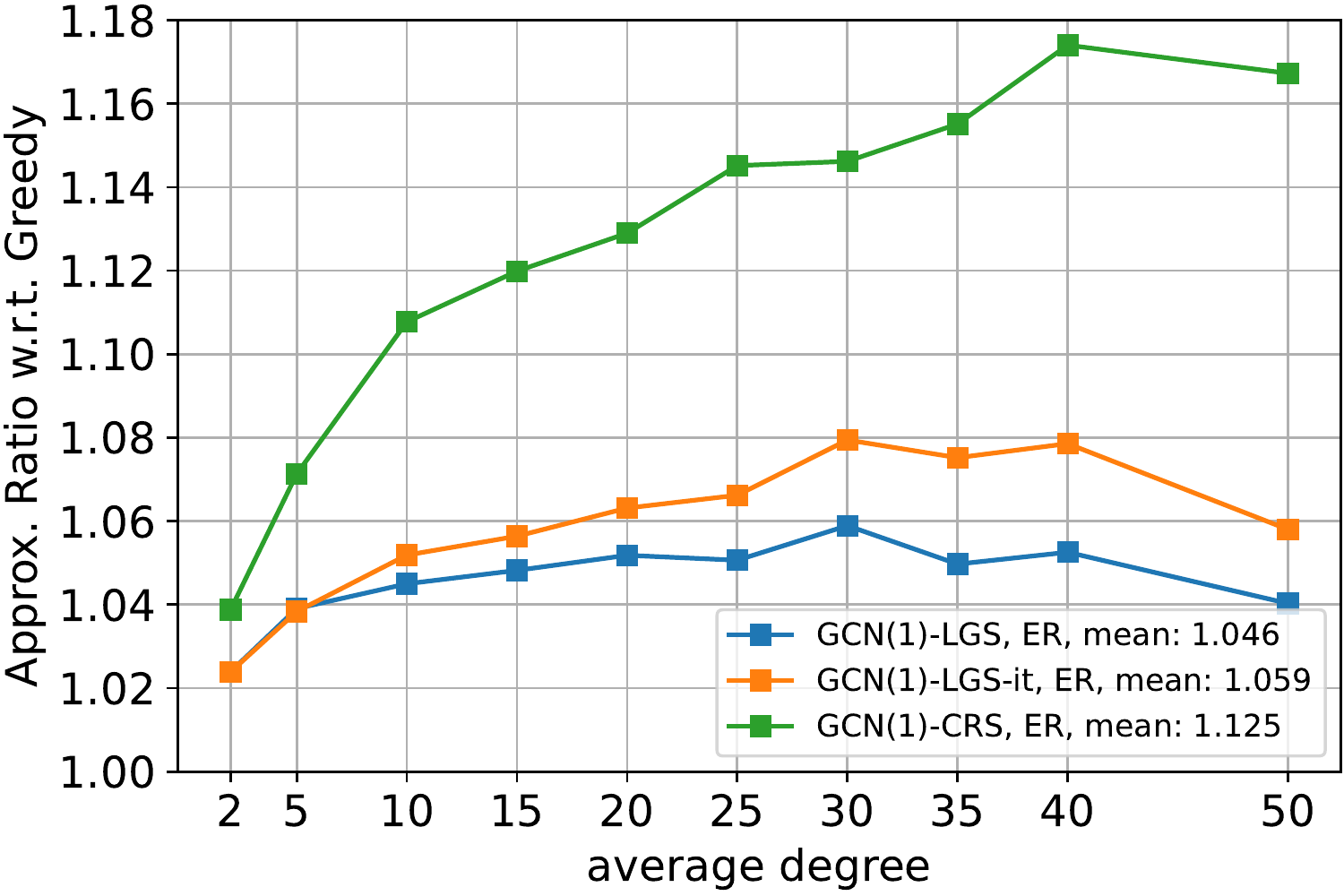}
    \label{fig:er:size}
    }\\
    \subfloat[]{
    \includegraphics[width=0.9\linewidth]{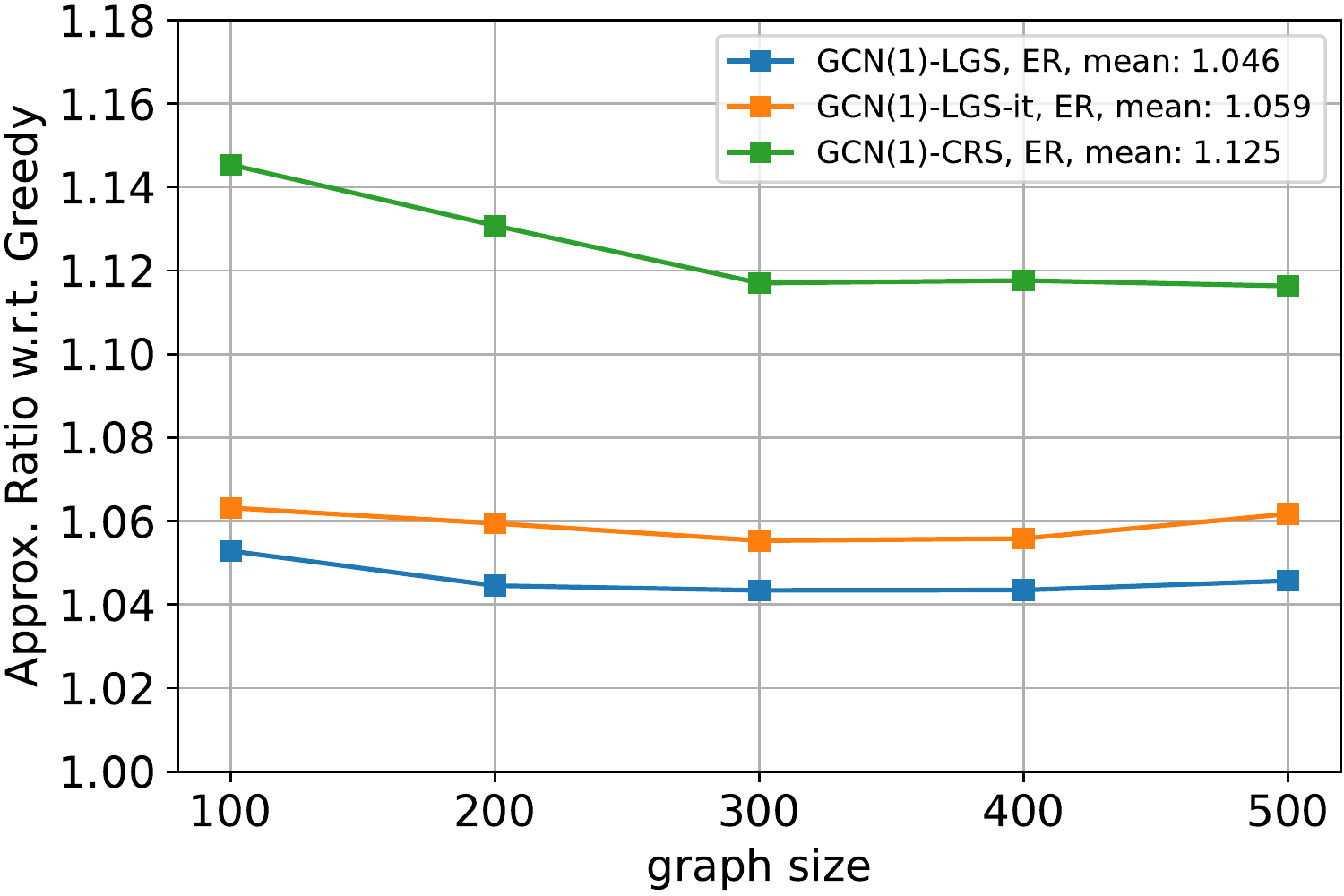}
    \label{fig:er:degree}
    }
    \caption{Approximation ratios of GCN-based MWIS solvers w.r.t. vanilla LGS, on a set of 500 ER graphs as a function of (a)~average degree and (b)~graph size, where the GCN is trained on ER graphs of average degree $\leq 12.5$ and size $|\ccalV|\leq300$. }\label{fig:er}
    \vspace{-0.2in}
\end{figure}%

\begin{figure}[t]
    \vspace{-0.1in}
    \centering
    \subfloat[]{
    \includegraphics[width=0.9\linewidth]{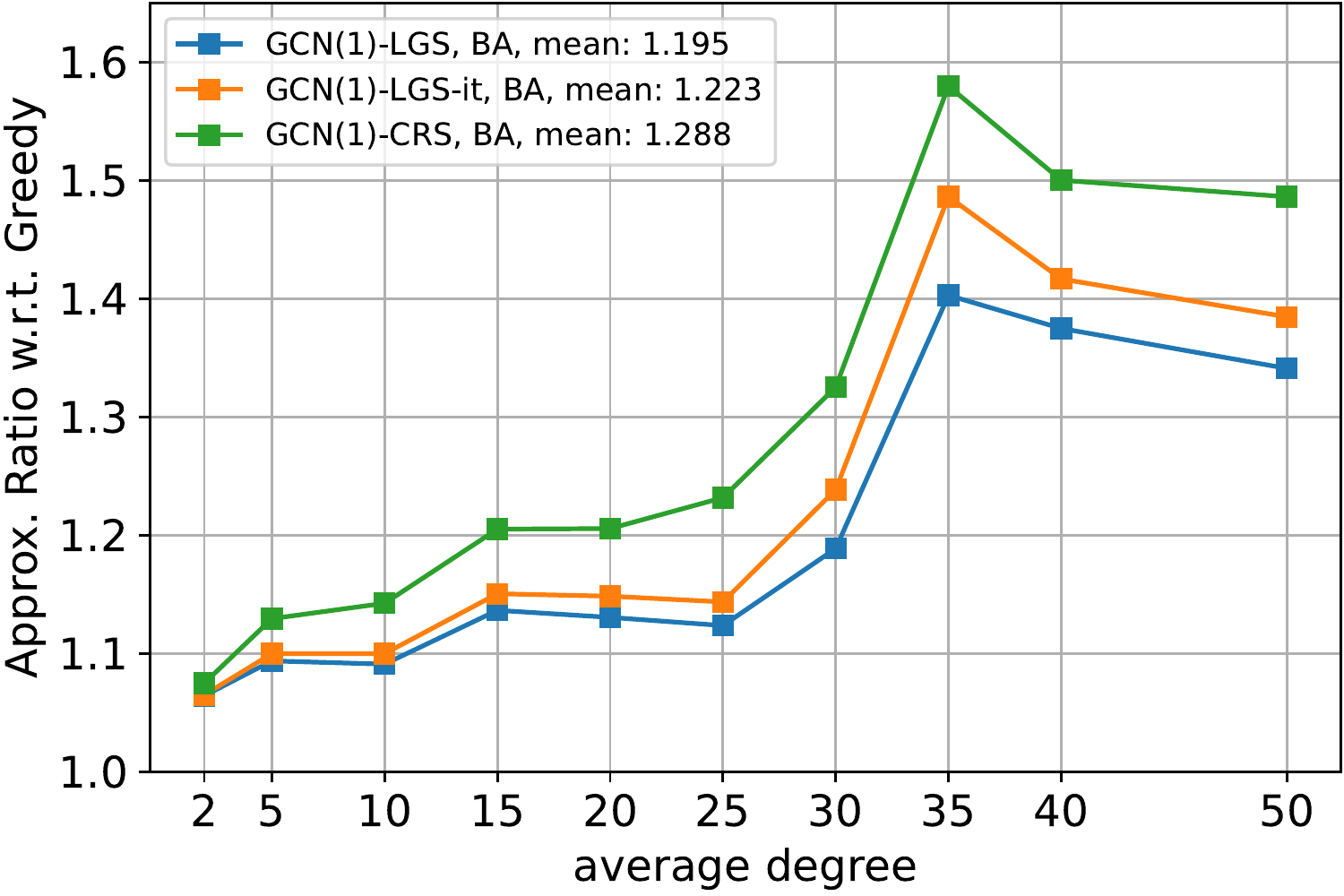}
    \label{fig:ba:size}
    }\\
    \subfloat[]{
    \includegraphics[width=0.9\linewidth]{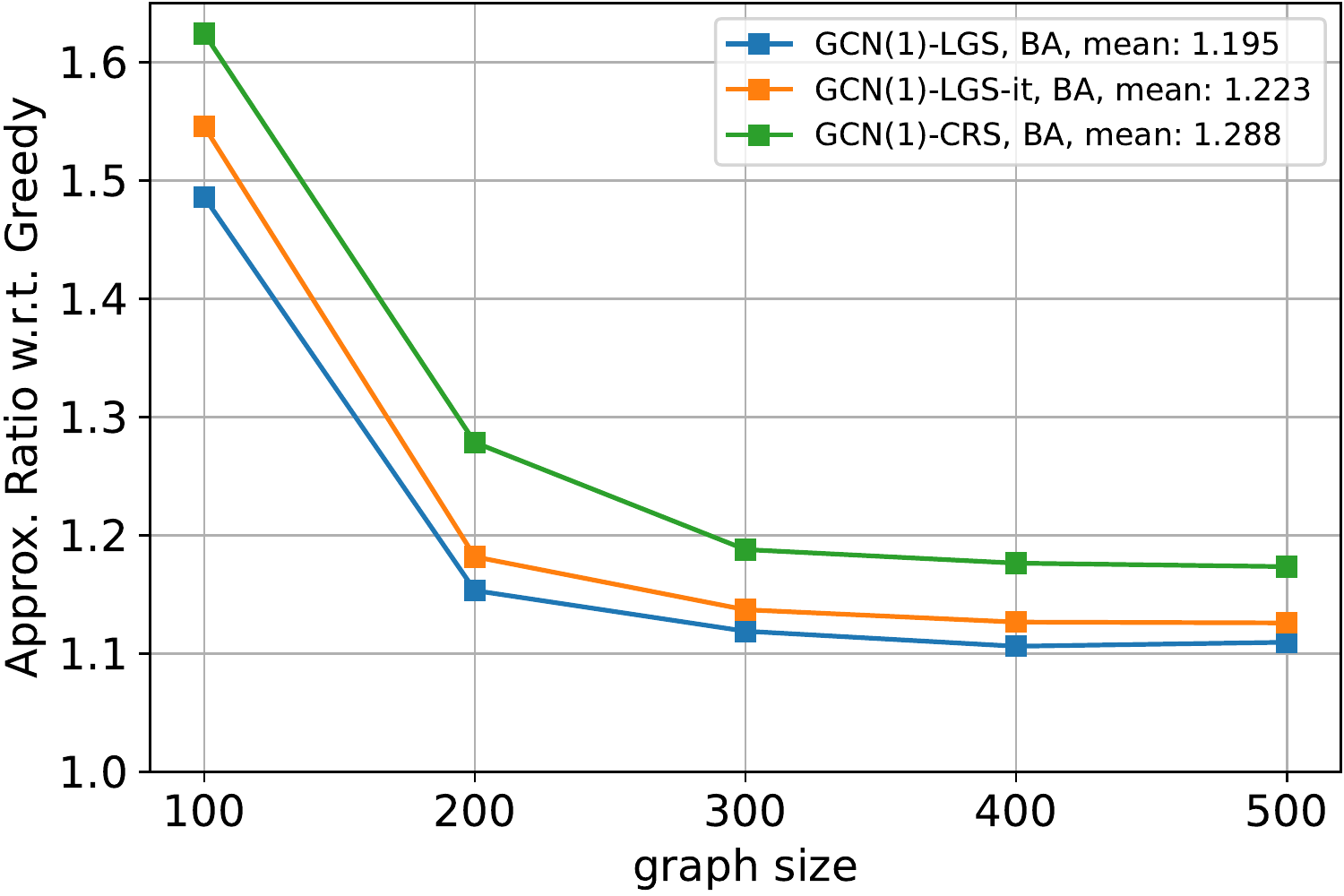}
    \label{fig:ba:degree}
    }
    \caption{Approximation ratios of GCN-based MWIS solvers w.r.t. vanilla LGS, on a set of 500 BA graphs as a function of (a)~average degree and (b)~graph size, where the GCN is trained on BA graphs of average degree $\leq 12.5$ and size $|\ccalV|\leq300$. }\label{fig:ba}
    \vspace{-0.2in}
\end{figure}%

\subsection{Generalizability to larger and denser graphs}

In practice, the size of the wireless networks could vary from time to time. It is important for GCN-based MWIS solvers to be able to generalize to graphs that are larger and/or denser than the graphs they are trained on.

We test the GCN(1)-CRS, GCN(1)-LGS, and GCN(1)-LGS-it respectively trained on the two training sets of ER and BA graphs specified in Section~VII of \cite{zhao2022twc}, on a set of 500 ER graphs (Figs.~\ref{fig:er}) and a set of 500 BA graphs (Figs.~\ref{fig:ba}).
These two test datasets have graphs with average degree of up to 50, and size of up to 500. 
The approximation ratios of the three GCN-based heuristics w.r.t. the local greedy solver, as functions of average degree and graph size, are presented in Figs.~\ref{fig:er} (for ER graphs) and Figs.~\ref{fig:ba} (for BA graphs).
The results show that with GCN trained on graphs of average degree $\leq 12.5$ and size $|\ccalV|\leq 300$, our proposed centralized and distributed solvers can still maintain their advantages over the baseline greedy solver, showing good generalizability towards larger and denser graphs.

\bibliographystyle{IEEEtran}
\bibliography{strings,refs}